\documentstyle[12pt,epsfig]{article}

\newlength{\dinwidth}
\newlength{\dinmargin}
\def\lapproxeq{\lower .7ex\hbox{$\;\stackrel{\textstyle
<}{\sim}\;$}}
\def\gapproxeq{\lower .7ex\hbox{$\;\stackrel{\textstyle
>}{\sim}\;$}}
\def\gtrsim{ \;\raisebox{-.7ex}{$\stackrel{\textstyle
>}{\sim}$}\; }
\def\lesim{ \;\raisebox{-.7ex}{$\stackrel{\textstyle
<}{\sim}$}\; }

\def\be{\begin{equation}}
\def\ee{\end{equation}}
\def\bea{\begin{eqnarray}}
\def\eea{\end{eqnarray}}

\def\bb{b\bar{b}}
\def\cc{c\bar{c}}
\def\qq{q\bar{q}}
\def\ra{ \rightarrow }
\def\whs{\widehat{\sigma}}
\def\GeV{{\rm GeV}}

\begin{document}
\begin{flushright}
IPPP/03/05 \\
DCPT/03/10 \\
DESY 02--227 \\
14 May 2003 \\
\end{flushright}

\vspace*{2cm}

\begin{center}
{\Large \bf Prompt neutrinos from atmospheric $\cc$ and $\bb$ production\\
and the gluon at very small
\renewcommand{\thefootnote}{\fnsymbol{footnote}}$x\,$\footnote[2]{To
appear in the special issue of Acta Physica Polonica to celebrate
the 65th Birthday of Professor Jan
Kwieci\'nski.}\renewcommand{\thefootnote}{\arabic{footnote}}\setcounter{footnote}{0}
\rule{0mm}{2.5ex}}

\vspace*{1cm}
\textsc{A.D. Martin$^a$, M.G. Ryskin$^{a,b}$ and A.M. Sta\'sto$^{c,d}$} \\

\vspace*{0.5cm} $^a$ Institute for Particle Physics Phenomenology,
University of Durham, DH1 3LE, UK \\
$^b$ Petersburg Nuclear Physics Institute, Gatchina,
St.~Petersburg, 188300, Russia \\
$^c$ Theory Division, DESY, D22603 Hamburg, Germany \\
$^d$ H. Niewodnicza\'nski Institute of Nuclear Physics, 31-342
Krakow, Poland
\end{center}

\vspace*{1cm}

\begin{abstract}
We improve the accuracy of the extrapolation of the gluon distribution of the proton to very small $x$, and show
that the charm production cross section, needed to calculate the cosmic ray-induced `atmospheric' flux of
ultrahigh  energy prompt $\nu_\mu$ and $\nu_\tau$ neutrinos, may be predicted within perturbative QCD to within
about a factor of three. We follow the sequence of interactions and decays in order to calculate the neutrino
fluxes as a function of energy up to $10^9\ \GeV$. We also compute the prompt $\nu_\tau$ flux from $\bb$
production, hadronization and decay. New cosmic sources of neutrinos will be indicated if more prompt neutrinos
are observed than predicted. If fewer neutrinos are observed than predicted, then constraints will be imposed on
the nuclear composition of cosmic rays. The advantages of studying $\nu_\tau$ neutrinos are emphasized. We provide
a simple parameterization of the prediction for the inclusive cross section for $c$ quark production in high
energy proton--air collisions.
\end{abstract}

\section{Introduction}

Very high energy `cosmic' neutrinos with energies in excess of
about 10~TeV offer a unique source of valuable information about
energetic events far away in the Universe; see, for example, the
reviews in Refs.~\cite{NUREV,UNU}. This has led to the development
of neutrino telescopes, which use photo-multiplier tubes to detect
the Cerenkov radiation emitted from the charged leptons produced
in charged-current neutrino interactions in a large volume of
water or ice, deep underground; see, for example, the reviews in
Refs.~\cite{UNU,SPIER}. If we consider muon neutrinos, then up to
about 100~TeV the spectrum is dominated by atmospheric neutrinos
from the decays of pions or kaons produced by cosmic ray
interactions in the Earth's atmosphere. At higher energies the
increased lifetime of these mesons means that they interact before
they have the opportunity to decay. Above 500~TeV the decays of
the much shorter-lived charmed particles become the dominant
source of atmospheric muon neutrinos. These are known as `prompt'
neutrinos\footnote{For tau neutrinos, we will see that prompt
production dominates at about 10~TeV and above.}. Their energy
dependence follows the original cosmic ray spectrum, while the
spectrum of `conventional' atmospheric neutrinos falls off by an
extra power of the energy because of the competition between the
decay and interaction of the parent meson. It is clearly essential
to quantify the flux of `prompt' neutrinos as accurately as
possible, since they provide the background to the sought-after
`cosmic' neutrinos.

There exist many models of the `prompt' neutrino flux in the
relevant $10^4$--$10^9$~GeV energy range, which yield predictions
which differ by more than two orders of magnitude. Some of the
models are purely phenomenological and have arbitrary continuation
to high energy from the domain constrained by accelerator data.
However, it has been noted that, with the inclusion of the
next-to-leading order (NLO) contributions~\cite{NDE}, perturbative
QCD gives a satisfactory description of the observed features of
the available accelerator data on charm production, see for
example~\cite{APPEL}--\cite{GGV3}. Moreover, the simplified LO
calculation reproduces the same behaviour when multiplied by an
overall $K$ factor, $K\simeq2.3$, to account for the NLO
contribution. Several authors have therefore used perturbative QCD
to predict the prompt neutrino flux~\cite{TIG}--\cite{GGV3}.

The perturbative approach, however, faces the same problem of
extrapolation to high energies. The LO diagram for forward high
energy charm production is shown in Fig.~1. The cross section may
be written in terms of the Feynman variable $x_F = p_L/p_L^{\rm
max}$, where $p_L$ is the longitudinal momentum of the charm
quark; at high energies $x_F\simeq E_c/E$, where $E$ is the
incident proton energy and $E_c$ is the energy of the charm quark.
Using the notation of Fig.~1, we have
\begin{equation}
\frac{d\sigma}{dx_F}(pp\ra c+\dots)\ =\ \int\,dx_1\,dx_2\,dz\
g(x_1,\mu_F^2)\:\frac{d\bar\sigma_{gg\ra\cc}}{dz}\;g(x_2,\mu_F^2)\,\delta(zx_1-x_F),
\label{eq:dsigmabydx_F}
\end{equation}
where $d\bar\sigma/dz = \bar s\,d\bar\sigma/d\bar t$ with
$z=(m_c^2-\bar t)/\bar s$, and where $g(x)$ is the gluon density
of the proton. The Mandelstam variables $\bar s$ and $\bar t$
refer to the $gg\ra \cc$ subprocess.\footnote{Throughout we take
the mass of the charm quark to be $m_c=1.25$~GeV, following
Ref.~\cite{GGV1}. We know that this value, taken together with the
NLO contribution ($K\simeq 2.3$), gives a good description of all
the available fixed-target data for $\cc$, or rather D meson,
production (which are in the region $E\sim250$~GeV)~\cite{GGV1}.}
The problem is that in the high energy domain of interest we
sample the gluon density at very small $x_2$; to be specific
$x_2\simeq M_{\cc}^2/2x_Fs\sim10^{-9}$--$10^{-4}$, where $\sqrt s$
is the total $pp$ c.m. energy. There are no data which determine
the gluon for $x\lapproxeq10^{-4}$, and, as a rule, parton
distributions for $x<10^{-5}$ are not available. For example,
Ref.~\cite{GGV3} shows a range of predictions for the prompt flux
neutrino obtained by continuing the gluon distribution below
$x=10^{-5}$ using the power law dependence $xg\sim x^{-\lambda}$
with $\lambda$ in the range 0--0.5. Of course, the prediction
depends crucially on the value taken for $\lambda$, and at the
highest neutrino energy shown, $10^9$~GeV, the rates span about
two orders of magnitude.

\begin{figure}[h]
\begin{center}
\epsfig{figure=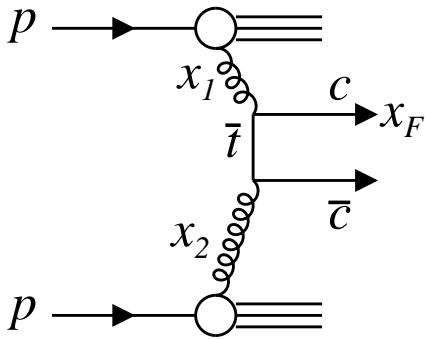,height=2in} \caption{The lowest-order
diagram for $\cc$ production in high energy $pp$ collisions. The
`small $x$' gluon has typical values $x_2\simeq M_{\cc}^2/2sx_F$,
where $x_F\sim 0.1$.}
\end{center}\end{figure}

The goal of the present paper is to diminish the uncertainty in
the predictions of the prompt neutrino flux. The major problem is
to obtain the most reliable method of extrapolation based on the
present understanding of the small $x$ regime. In order to do this
we begin, in Section~\ref{sec:gluonatsmallx}, by comparing
different physically-motivated extrapolations of the gluon to very
small $x$:
\begin{itemize}
\item[(i)] DGLAP gluon with a double leading log (DLL)
extrapolation,
\item[(ii)] unified DGLAP/BFKL gluon \cite{KMS} with
$x^{-\lambda}$ extrapolation,
\item[(iii)] extrapolation with saturation effects included.
\end{itemize}

In Section~\ref{sec:predictions} we compare the predictions for
the $x_F$ distribution  of charm quarks produced in high energy
$pp$ collisions ($E\sim10^5$ to $10^9$~GeV) using the three models
for extrapolating the gluon into the small $x$ regime, typically
$x\sim10^{-4}$ to $10^{-8}$. We argue that the extrapolation based
on model~(iii) is the most reliable, and so for the remainder of
the paper we show results for this gluon. To determine the prompt
neutrino flux we need to extend the calculation to high energy
proton--air collisions. This is done in
Section~\ref{sec:ccbarproduction}, where we also consider the
uncertainties associated with the extrapolation based on
model~(iii).

In Section~\ref{sec:promptneutrinos} we describe the formalism
that we shall use to calculate the prompt lepton spectra. Starting
from the production of $\cc$ pairs from the incoming cosmic ray
flux, we allow for their fragmentation into charmed ($D^\pm$,
$D^0$, $D_s$ and $\Lambda_c$) hadrons, and for their subsequent
semileptonic decays. We include the effects of the lifetime of the
charmed hadrons and, also, for the lifetime and decay modes of the
$\tau$ lepton in the $D_s\ra\tau\nu_\tau$ decays. The results of
the calculation of the prompt $\nu_\mu$ and $\nu_\tau$ fluxes are
presented in Section~\ref{sec:promptnufluxes}, and compared with
the `conventional' atmospheric fluxes. These latter fluxes arise
from $\pi$, $K \dots$ decays and $\nu_\mu-\nu_\tau$ oscillations
respectively. We find that the prompt $\nu_\tau$ spectrum for
$E>10^4$~GeV lies much above its conventional atmospheric
background, whereas for the prompt $\nu_\mu$ spectrum this is only
achieved for $E>10^6$~GeV. The origins of the prompt $\nu_\tau$
flux are the $D_s\ra\tau\nu_\tau$ decays which occur with a
branching fraction of $6.4\pm 1.5\%$~\cite{RPP2002}. It is
relevant to note that high energy $\nu_\tau$'s, unlike
$\nu_\mu$'s, are not depleted in number by absorption in the
Earth. They will always penetrate the Earth due to the
$\nu_\tau\ra\tau\ra\nu_\tau\dots$ regeneration
sequence~\cite{NUTAU}. This is clearly crucial for upgoing high
energy neutrinos through the Earth and can be important for
horizontal neutrinos (particularly of high energy) entering a
detector deep underground.

At first sight, it appears that the prompt neutrino flux from
$\bb$ production will, relative to the flux of $\cc$ origin, be
suppressed, first, by a factor of order $m_c^2/m_b^2$ and, second,
by the fact that the gluon is sampled at larger $x$. However, for
the prompt $\nu_\tau$ flux of $\bb$ origin, the suppression is
partly compensated by the existence of significant $\tau\nu_\tau$
semileptonic decays of all the beauty hadrons ($B^\pm$, $B^0$,
$B_s$ and $\Lambda_b$), in contrast to just the
$D_s\ra\tau\nu_\tau$ decays for charm. We calculate the prompt
flux arising from $\bb$ production, fragmentation and decay in
Section~\ref{sec:prompt}. Finally, in the concluding section, we
discuss the implications of our results for neutrino astronomy and
cosmic ray physics. Also, there, we summarize the uncertainties in
the predictions of the prompt neutrino fluxes.

\section{The gluon at small $x$ and high energy $\cc$
production} \label{sec:gluonatsmallx}

As mentioned above, one possible method of extrapolating the gluon
into the $x<10^{-5}$ regime is to resum the leading $\alpha_S \ln
Q^2\ln 1/x$ terms within the DGLAP framework, which leads to a
small $x$ behaviour\footnote{In practice, we use the DLL
continuation to $x<10^{-5}$ in the form
$$g(x,Q^2)\ \ =\ \
g(x=10^{-5},Q^2)\,\exp\left(\sqrt{\frac{16N_C}{b}\ln\frac{\alpha_S(Q)}{\alpha_S(Q_0)}\ln\frac{x}{x_0}}
\;-\;
\sqrt{\frac{16N_C}{b}\ln\frac{\alpha_S(Q)}{\alpha_S(Q_0)}\ln\frac{10^{-5}}{x_0}}\right),$$
where the LO coupling $\alpha_S(Q) =
4\pi/(b\log(Q^2/\Lambda^2_{\rm QCD}))$ with $n_f=4$ and $b=25/3$.
We take $Q_0^2=1$~GeV$^2$ and $x_0=0.25$.  We use MRST2001
partons~\cite{MRST2001} with $\Lambda_{\rm QCD} = 220$~MeV. The
reliability of this form of DLL extrapolation has been checked
using GRV partons~\cite{GRV} which are tabulated down to
$x=10^{-9}$.\vspace{1ex}}
\begin{equation} xg(x,Q^2)\ \simeq\
x_0g(x_0,Q_0^2)\,\exp\left(\sqrt{\frac{16N_C}{b}\ln\frac{\alpha_S(Q)}{\alpha_S(Q_0)}\ln\frac{x}{x_0}}\right).
\label{eq:xg}\end{equation} We denote this extrapolation by MRST
on the Figures below.

As far as we fix the scale $Q^2$ and extrapolate to much smaller
$x$, the leading contribution comes from $\alpha_S\ln1/x$ terms,
which, at leading order, are resummed by the BFKL
equation~\cite{LLBFKL}. Therefore a more reliable extrapolation is
obtained by solving a unified DGLAP/BFKL equation~\cite{KMS} for
the gluon. This equation is written in terms of the gluon,
unintegrated over its transverse momentum, which should be used
with the off-mass shell\footnote{That is we use the replacement in
(\ref{eq:dsigmabydx_F})
$$xg(x,\mu_F^2)\sigma_{gg\ra\cc}\ \ \ra\ \ xg(x,Q_0^2)\sigma_{\rm
on}\: + \: \int_{Q_0^2=1\ {\rm GeV}}^\infty\,
f_g(x,Q_t^2)\sigma_{\rm off}\,dQ_t^2$$ with $x=x_2$; where $f_g$
is the unintegrated gluon density as defined in \cite{KMS}, and
where $\sigma_{\rm on,\ off}$ are the $gg\ra\cc$ cross sections
with the small $x$ gluon on-, off-mass-shell. The large $x_1$
gluon, which is clearly in the DGLAP regime, is always taken to be
on-mass-shell.} matrix element for the hard $gg\ra\cc$ subprocess
amplitude~\cite{CCH}. In this way the result embodies the main
part of the NLO DGLAP correction. Besides this, the unified
equation includes a kinematic constraint (or consistency
condition)~\cite{KC} which accounts for the major part of the
higher-order corrections to the BFKL evolution.\footnote{Explicit
expressions for the next-to-leading log BFKL terms can be found
in~\cite{NNLBFKL}.} Indeed, the power behaviour generated
corresponds to $\lambda\simeq0.3$ which is much less than the LO
BFKL behaviour $x^{-\omega_0}$ with $\omega_0 = 12\alpha_S\ln
2/\pi$. Moreover, the charm production cross section
$d\sigma/dx_F$ calculated in terms of the KMS unintegrated
gluons~\cite{KMS} is found to coincide, within 10\% accuracy, with
the prediction obtained with conventional DGLAP
gluons~\cite{MRST2001} for $10^4$--$10^5$~GeV laboratory energies,
corresponding to $x=10^{-3}$--$10^{-4}$ for which deep inelastic
accelerator data exist. Note that, in this comparison, the
prediction based on conventional partons was calculated at LO and
multiplied by $K=2.3$, which, according to Ref.~\cite{GGV1},
accounts well for the NLO corrections. On the other hand the
unified DGLAP/BFKL prescription already incorporates the main NLO
effect at small $x$, and so no $K$ factor is present in this $x$
domain. We denote the results obtained using this gluon by KMS on
the Figures below.

At $10^5$~GeV these predictions should be reliable. However, as we
proceed to higher energies, we sample gluons in smaller and
smaller $x$ regimes with increasing gluon density, and so we must
account for the effect of saturation. To study this effect we
start with the Golec-Biernat,~W\"{u}sthoff~(GBW) model of deep
inelastic scattering (DIS)~\cite{GBW1} (and diffractive
DIS~\cite{GBW2}). Let us outline the basis of the model, as
applied to $\qq$ production in DIS. The production of $\qq$ pairs
is described by the probability of the formation of the pair by
the initial photon multiplied by the cross section for the
$\qq$--proton interaction, $\whs$. The first stage is given by the
effective photon wave functions $\Psi_{T,L}$ (for transverse,
longitudinal polarisations), which depend on the momentum fraction
$z$ carried by the quark and the transverse separation $\vec{r}$
between the $q$ and $\bar{q}$. The deep inelastic cross sections
have the form~\cite{NZ}
\begin{equation}
\sigma_{T,L}(x,Q^2)\ =\ \int
d^2\,r\int_0^1dz\;\sum_q\left|\Psi_{T,L}^q(\vec{r},z,Q^2)\right|^2\:\whs(x,\vec{r}).
\label{eq:sigma_T,L}
\end{equation}
For small $r$, the dipole cross section $\whs$ is proportional to
$r^2$. To allow for multiple interactions, Golec-Biernat and
W\"{u}sthoff~\cite{GBW1} parameterize $\whs$ by the form
\begin{equation}
\whs(x,r)\ =\
\sigma_0\left(1-\exp\left(-\frac{r^2}{4R_0^2(x)}\right)\right),
\label{eq:sigmahat}
\end{equation}
with an $x$-dependent saturation radius
\begin{equation}
R_0(x)\ =\ \frac{1}{Q_0}\left(\frac{x}{x_0}\right)^{\lambda/2}.
\label{eq:R_0}
\end{equation}

The parameterization is a simplified version of the well-known
Glauber expression for, say, describing the multiple interactions
of a pion passing through a nucleus
\begin{equation}
\sigma_{\pi A}\ =\ \int d^2b_t\:\left[1- \exp\left(-\sigma_{\pi
N}T(b_t)\right)\right]. \label{eq:sigma_piA}
\end{equation}
The integral $\int d^2b_t$ gives the nuclear area $\pi R^2_A$,
which is replaced by $\sigma_0$ in (\ref{eq:sigmahat}), and the
mean nucleon density $\langle\, T\, \rangle$ is parameterized by
$1/R_0(x)^2$, modulo normalisation. $\sigma_{\pi N}$ is the
$\pi$--nucleon cross section, which is equivalent to the
$\qq$-dipole cross section in the GBW~model. That is, the exponent
$\sigma_{\pi N}T$ in (\ref{eq:sigma_piA}) is equivalent to
$(\pi^2r^2\alpha_S/3)xg/\sigma_0$, where the gluon density
$xg/\sigma_0$ plays the role of the mean nucleon density
$\langle\, T\, \rangle$, and where the factor in brackets plays
the role of $\sigma_{\pi N}$. It is because the gluon density
grows as $x$ decreases that we have an $x$ dependence in the
argument of the exponential in (\ref{eq:sigmahat}).

The GBW model has recently been realised~\cite{GBevol} in terms of
the gluon density, including the DGLAP $\ln Q^2$ evolution of
$g(x,Q^2)$. Actually in this improved form it should be
considered, not as a model, but as a complete perturbative
calculation, which in addition to the conventional LO collinear
approach also accounts for the rescattering of the incoming $\qq$
pair.

The power of $x$ in (\ref{eq:R_0}) reflects the power growth of
the gluon density in the small $x$ region. The parameters
$\sigma_0,x_0$ and $\lambda$ were fitted to describe the small $x$
DIS data~\cite{GBW1}. It was shown that, up to
$Q^2\simeq20$~GeV$^2$, a good description can be achieved, even
without accounting for DGLAP evolution. Interestingly, the value
of the power, $\lambda=0.28$, turns out to be close to the value
found in Ref.~\cite{KMS}.

So far we have considered absorption for DIS. Here we are
interested in $gg\ra\cc$, and not $\gamma g\ra\cc$. It is
therefore necessary to multiply the $\gamma g\ra\cc$ cross section
by
\be \label{eq:multiplicationfactor}
\frac{1}{6}\left[1\,-\,\frac{9}{4}\,z(1-z)\right]\frac{\alpha_S}{\alpha\,e_c^2}
\ee
where the first factor is due to colour, and the second term in
square brackets accounts for $gg\ra g\ra\cc$ production (see, for
example, \cite{COMBMAX}).

Note that the approach of Golec-Biernat and W\"{u}sthoff includes
only the rescattering of the $\cc$ pair and neglects the enhanced
Reggeon diagrams which account for the rescattering of the more
complicated Fock components of the photon (gluon) like $\cc g$,
$\cc gg$, etc. These extra components have a larger absorptive
cross section. In other words, when the gluon density becomes
sufficiently large, we must allow for $gg$ recombination, which
diminishes the rate of $\cc$ production. From this $t$ channel
viewpoint, the absorption is driven by the triple-Pomeron
interaction. With the help of the Balitskii--Kovchegov
equation~\cite{BALIT,KOV}, we may sum up the resulting fan
diagrams (formed from different networks of Pomeron--Pomeron
recombinations into single Pomeron exchanges). This effect has
been studied numerically in recent papers~\cite{GMS,BRAUN,LGLM}.
However, the approach is based on LO~BFKL, and does not account
for the NLO corrections, which are known to be large. In this case
we cannot simply rescale the LO prediction by taking a lower
Pomeron intercept,~$\omega_0$. The problem is that the
triple-Pomeron vertex is not known at NLO.

There are reasons, both from phenomenology~\cite{KAID} and from
perturbative evolution~\cite{BRV}, to believe that the
triple-Pomeron coupling is small. Nevertheless, at very high
energy, we expect the resulting absorption to be stronger. From
this point of view we may regard the prediction based on the GBW
model as the upper limit for $\cc$ production. Later, for a more
realistic estimate of the cross section for $\cc$ production, we
take account of the triple-Pomeron vertex by
replacing~(\ref{eq:R_0}) by\footnote{Such a form is motivated by
the results of Ref.~\cite{AKLR}. Recall that the combinatorial
factor which corresponds to the fan diagrams is $N!/N!\ra 1$.
Thus, contrary to the eikonal form, where we have
$\sum_N(-g)^N/N!=\exp(-g)$, here we deal with a geometrical series
type of expression $\sum_N(-g)^N=1/(1+g)$. Therefore we choose
form~(\ref{eq:R_0^2}) with the constant $c$ as evaluated in
\cite{KAID,BRV}.}
\be \label{eq:R_0^2} R_0^2\ =\
\frac{1}{Q_0^2}\left(c+(x/x_0)^\lambda\right) \ee
for $x<10^{-3}-10^{-4}$ with $c\simeq0.05$--$0.2$. This is to
protect $R_0$ becoming too small for small $x$.

The only reason why the above upper limit may be exceeded arises
because the GBW saturation model \cite{GBW1} was formulated at
fixed impact parameter, and so does not allow for the growth of
the proton radius $R_p$ with increasing energy. The radius $R_p$
can be determined from the slope $B$ of the elastic $pp$ cross
section,
\be \label{eq:B} B=B_0+2\alpha^\prime\ln E, \ee
where $\alpha^\prime$ is the slope of the Pomeron trajectory, and
$E$ is the proton energy in the laboratory frame. Indeed, for a
large-size dipole the GBW model saturates at
$\sigma=\sigma_0=29$~mb, whereas the normal soft hadronic cross
sections, which should be equivalent to large-size dipoles,
continue to grow logarithmically with energy. From a physical
point of view, the normalisation $\sigma_0$ in (\ref{eq:sigmahat})
is related to the proton area $\pi R_p^2$. Of course, we only have
the inequality $\sigma_0<\pi R_p^2\propto B$, since charm
production originates mainly from the centre of the proton.
However, since $\pi R_p^2$ grows with energy, a conservative {\em
upper limit} is obtained by multiplying the prediction obtained
from the GBW model by the factor $B(E)/B(E_0)$, with
$E_0\simeq10^4$~GeV, typical of the HERA domain where the
parameters of the model were tuned.

A {\em lower limit} to $\cc$ production may be obtained if we
assume a scaling behaviour for $dn_c/dx_F = (d\sigma(pp\ra
c+\dots)/dx)/\sigma_{\rm inel}$, where $\sigma_{\rm inel}$ is the
total inelastic $pp$ cross section -- that is if we assume that
$dn_c/dx_F$ is independent of energy. Hence the lower limit is
\begin{equation}
\frac{d\sigma(E)}{dx_F}\ =\
\frac{d\sigma(E_0)}{dx_F}\,\frac{\sigma_{\rm inel}(E)}{\sigma_{\rm
inel}(E_0)}\,, \label{eq:lowerlimit}
\end{equation}
normalised in the region $E\sim10^5$~GeV ($x\sim10^{-4}$) where
the parton distributions are known. To be more precise we should
replace $\sigma_{\rm inel}$ in (\ref{eq:lowerlimit}) by the cross
section corresponding to the Fock component of the proton wave
function which contains charm. However, the cross section for this
component will grow with energy faster than $\sigma_{\rm inel}$,
and thus (\ref{eq:lowerlimit}) may be regarded as an extreme lower
limit for the charm yield. We consider the Fock charm component to
be generated perturbatively. In principle, it would be possible to
have also a non-perturbative `intrinsic' charm
component~\cite{BROD}, although there is no firm experimental
evidence for its existence. Such an intrinsic charm cross section
would originate from the non-perturbative large-size domain,
controlled by $\sigma_{\rm inel}$, and hence its contribution
would become less important, with increasing energy, than the
perturbative cross section.

\section{Predictions for high energy $\cc$ production}
\label{sec:predictions}

In Fig.~2 we compare the predictions for the $x_F$ distribution of
charm quarks produced in $pp$ collisions, as given by the three
models for extrapolating the gluon\footnote{The gluons in
(\ref{eq:dsigmabydx_F}) are evaluated at a scale $\mu_F$ equal to
the transverse mass of the charm quark for the MRST and KMS
models; that is $\mu_F^2=m_c^2+p_{cT}^2$. For the GBW
extrapolation we take $\mu_F=\langle1/r\rangle$, where $r$ is the
separation between the $c$ and $\bar c$ quarks.} to small $x$. For
laboratory energies $E\sim10^3$--$10^5$~GeV we sample the gluons
in the $x$ region $10^{-2}$--$10^{-4}$ where the parton
distributions are known from global analyses. Hence, since each
model reproduces the same data, they give essentially the same
predictions for $\cc$ production. Recall that the LO DGLAP result,
based on MRST partons, was multiplied by a $K$ factor of 2.3. It
was shown~\cite{GGV1} that such a constant $K$ factor reproduces
well the NLO perturbative QCD prediction and gives a good
description of the available fixed-target data for $\cc$, or
rather $D$ meson, production for $E\sim250$~GeV. Recall that,
following~\cite{GGV1}, we take the mass of the charm quark to be
$m_c=1.25\ \GeV$. Although we use a constant $K$ factor, $K=2.3$,
we have confirmed that the use of the parameterization of the $K$
factor, $K(E_c,x_c)$, given in eq.~(3.4) of~\cite{PRS}, does not
appreciably alter any predictions.
\begin{figure}[h]
\begin{center}
\epsfig{figure=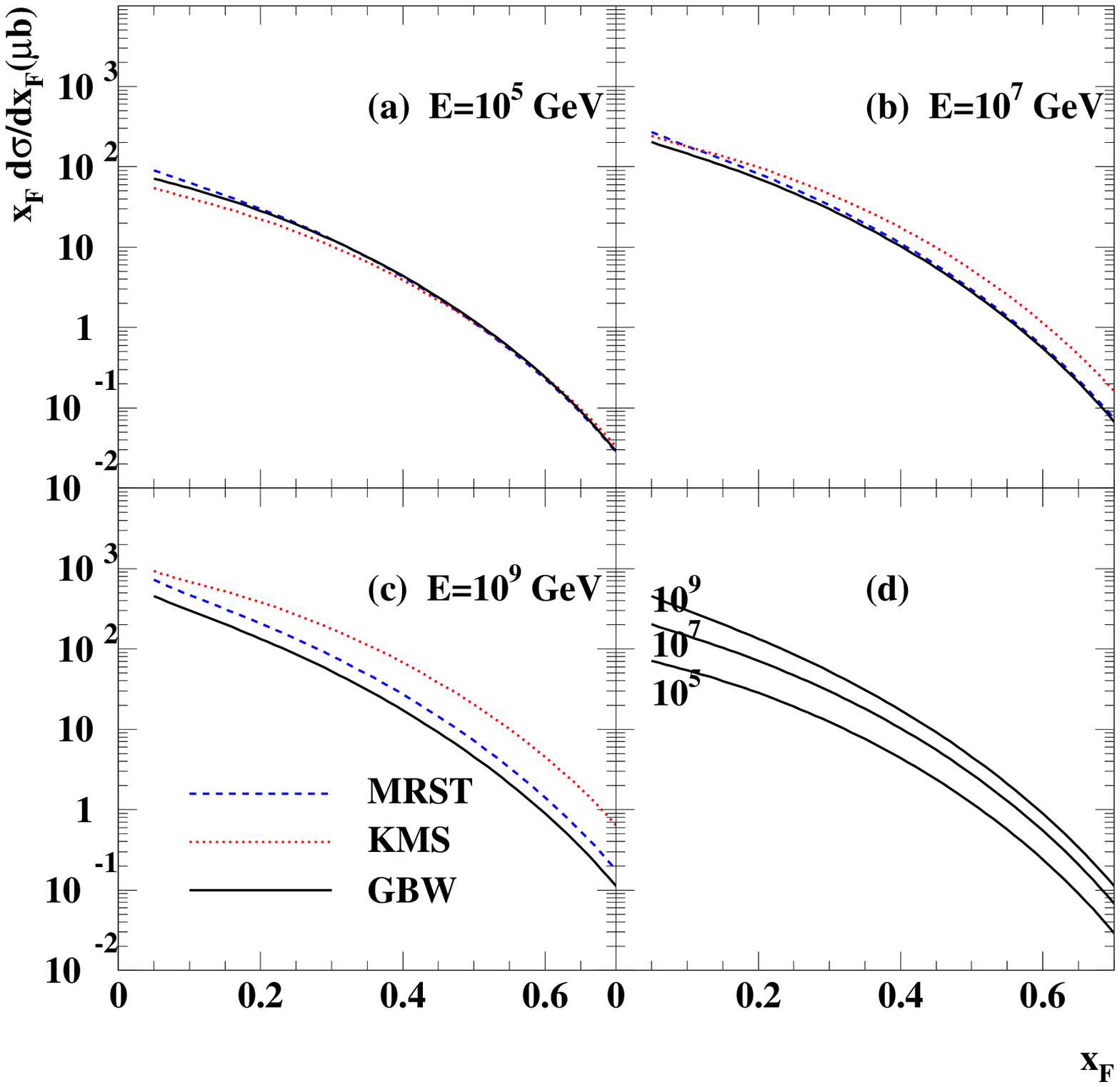,height=6in} \caption{The differential
cross section $x_Fd\sigma/dx_F$ for charm production in $pp$
collisions, (\ref{eq:dsigmabydx_F}), at three different laboratory
energies $E$, for three different ways of extrapolating the gluon
to very small $x$. The models are (i)~a double-leading-log DGLAP
extrapolation for $x<10^{-5}$ (MRST), (ii)~a unified DGLAP/BFKL
approach with an $x^{-\lambda}$ extrapolation for $x<10^{-7}$
(KMS), and (iii)~an extrapolation with saturation effects included
(GBW). Plot~(d) compares the GBW prediction at the three
energies.}
\end{center}\end{figure}

Up to $E\sim10^7$~GeV, the GBW saturation model practically
coincides with the DGLAP (MRST) prediction. For higher energies
the GBW cross section is lower due to absorptive effects. On the
contrary, the prediction based on KMS partons becomes higher, as
well as lower at the lower energies. The `unified' KMS evolution
includes the BFKL $\ln 1/x$ resummation and generates a power
growth, $x^{-\lambda}$, of the gluon density as we extrapolate to
small $x$. This evolution embodies a kinematic constraint (or
consistency condition) which accounts for a major part of the NLO
and higher-order BFKL effects. However, the power,
$\lambda\sim0.3$, is appreciable, and the growth exceeds the
double logarithmic DGLAP growth of the MRST result. Another
feature to note is that the shape of the $x_F$ distribution
becomes a little steeper with increasing energy, as seen in
Fig.~2(d), which shows the predictions of the GBW extrapolation
for three different energies.

On the other hand, at low energies, we see from Fig.~3 that the
KMS prediction falls away. Indeed, it is about a factor of two
below the GBW/MRST predictions for $E_c=10^3$~GeV. Here, we are
sampling the gluon at $x$ values above 0.01, where the $\ln 1/x$
resummation is not effective and where LO DGLAP evolution takes
over. To put it another way, it was observed in the KMS unified
BFKL/DGLAP approach~\cite{KMS}, that the cross section for an
off-shell gluon is enhanced by $\ln 1/x$ effects for
$x\lesim0.01$, whereas it rapidly tends to the LO on-shell DGLAP
formula as $x$ increases above this value. Hence in the low energy
regime a factor $K\sim2$ should be included in the KMS prediction,
with the factor dying away with increasing energy, as we enter the
BFKL regime.

A convenient way to summarize the relevant energy behaviour of the
$d\sigma^c/dx_F$ cross section is to plot the `$Z$ moment'
\cite{GAISSER} of the $x_F$ distribution, see for
example~\cite{COSTAckbk}. For high energies ($E>10^6$~GeV) the
incoming primary cosmic ray flux falls down as $E^{-(\gamma+1)}$
with $\gamma=2.02$. Therefore the charm flux is proportional to
the moment
\be \label{eq:momentZ_c} \sigma Z_c\ \equiv\
\int\frac{d\sigma^c}{dx}\,x^{2.02}\,dx. \ee
This moment is shown in Fig.~3,
\begin{figure}[h]
\begin{center}
\epsfig{figure=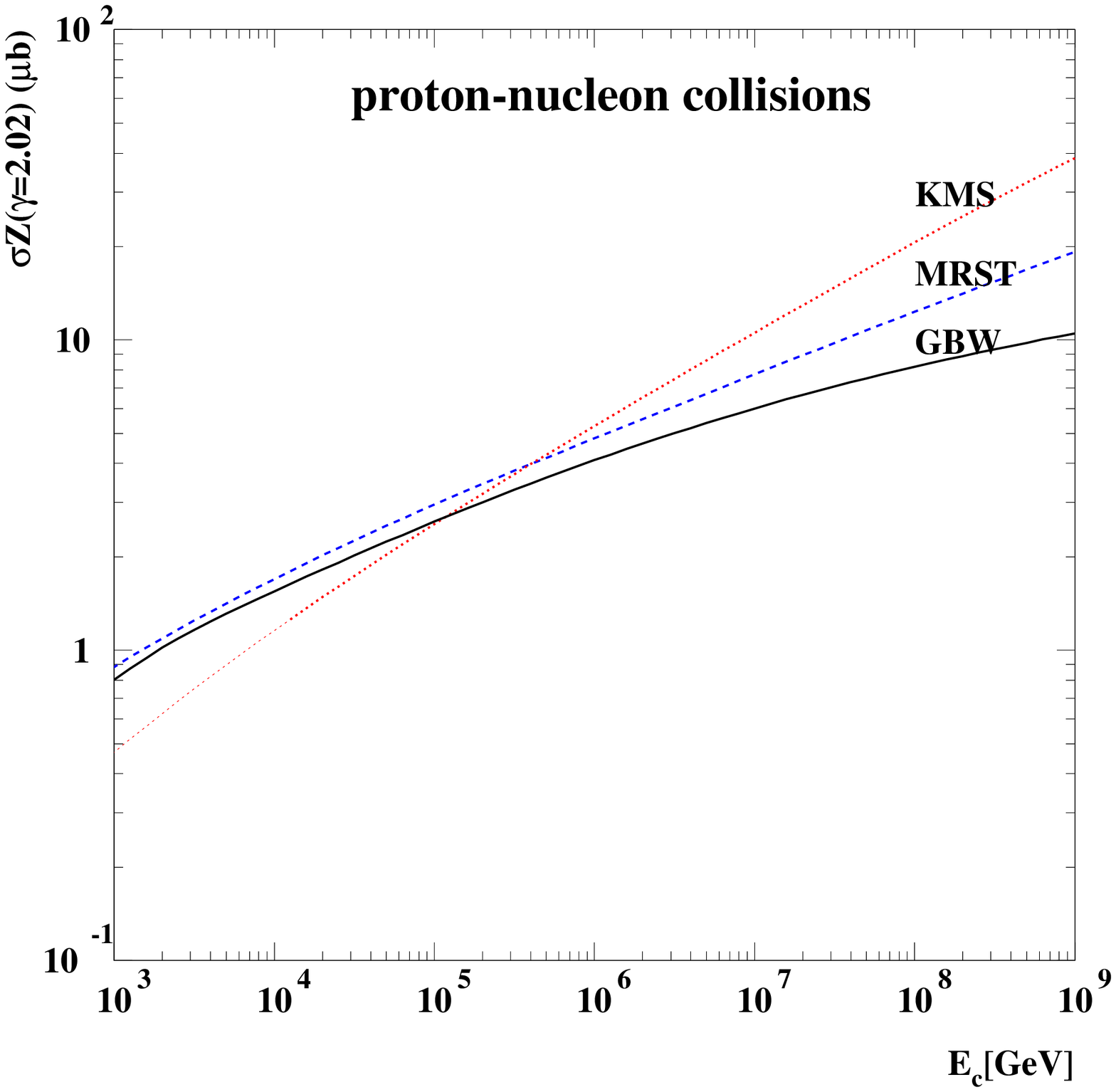,height=5.5in} \caption{The energy
dependence of the relevant `$Z$ moment', (\ref{eq:momentZ_c}), of
charm production in proton--nucleon collisions, as a function of
the energy of the produced charm quark. The models are as in
Fig.~2. The reason why the KMS prediction falls below the other
predictions at the smaller values of $E_c$ is explained in
Section~\ref{sec:predictions}. The $\gamma=2.02$ moments are shown
for illustration; in the calculation of the neutrino fluxes, the
differential cross section $x_F d\sigma /dx_F$ is convoluted with
the observed primary cosmic ray flux.}
\end{center}\end{figure}
where the difference between the saturation model and the other
two models becomes apparent for $E_c>10^6$~GeV. Note that here we
fix the energy $E_c$ of the outgoing $c$ quark, rather than that
of the incident proton which was used in Fig.~2.

Although the GBW model predicts the smallest cross section of the
three models, it should be considered as the {\em upper
limit}\footnote{Modulo a possible growth of the proton radius, see
(\ref{eq:B}).} for $\cc$ production as it only accounts for part
of the absorptive effects. For the reasons mentioned earlier, the
GBW model is more than a model -- rather it should be regarded as
a full leading-order QCD prediction with $\cc$ rescattering
effects included. Of course there is, in addition, absorption of
the gluons in the evolution process. It appears likely that the
consequent reduction of the $\cc$ cross section due to this
additional absorption may be partially compensated by the growth
of the proton radius with increasing energy. We investigate this,
and other effects, in the next Section; see Fig.~4 later.
Therefore, from now on, we will base our study on the GBW approach
and its variations.

\section{$\cc$ production in proton--air collisions}
\label{sec:ccbarproduction}

For a precise study we need the charm yield in $p$-nuclear (air)
collisions. An advantage of the GBW saturation model is that it
may be straightforwardly extended from $pp$ to $p$-nuclear
collisions. Within the GBW framework we may account for the
eikonal rescattering of the $\cc$ pair within the nucleus by
replacing $\whs$ in~(\ref{eq:sigmahat}) by
\be \label{eq:Asigmahatsum}
A\whs\sum_{n=1}^\infty\,\frac{(-(A-1)\whs/8\pi B_A)^{n-1}}{n\,n!}
\ee
where, for air, the mean atomic number $A=14.5$ and the slope
$B_A=\langle r^2\rangle/6=29\ \GeV^{-2}$. Note that in the
numerator we have $(A-1)$ and not $A$. In this way we exclude
rescattering on the nucleon where the $\cc$ pair is created. This
rescattering is controlled by a different slope ($\neq B_A$), and
is already accounted for in the cross section given in
(\ref{eq:sigmahat}). We have taken the root-mean-square nuclear
radius\footnote{For example, Ref.~\cite{ANCO} gives the r.m.s.
radius of oxygen as 2.7~fm. To obtain the result for air we take
the usual $r\propto A^{\frac{1}{3}}$ dependence.} $\sqrt{\langle
r^2\rangle} = 2.6$~fm, and assumed a gaussian distribution of the
nucleon density for light nuclei. Note that the replacement of
$\whs$ occurs before the integration over the $c$--$\bar{c}$
separation in (\ref{eq:sigma_T,L}). In summary, the inclusive
$\cc$ cross section for proton--air interactions is given by the
sum of proton--nucleon cross sections, with the only nuclear
effect being the enhanced absorption of the produced $\cc$ pair.
In the Appendix we give a simple parameterization which reproduces
the proton--air $\cc$ production cross section in the energy
interval $10^4<E<10^{11}~\GeV$ to within $\pm5\%$.

The $\gamma=2.02$ moment of charm production in high energy
proton--air collisions is shown in Fig.~4.
\begin{figure}[h]
\begin{center}
\epsfig{figure=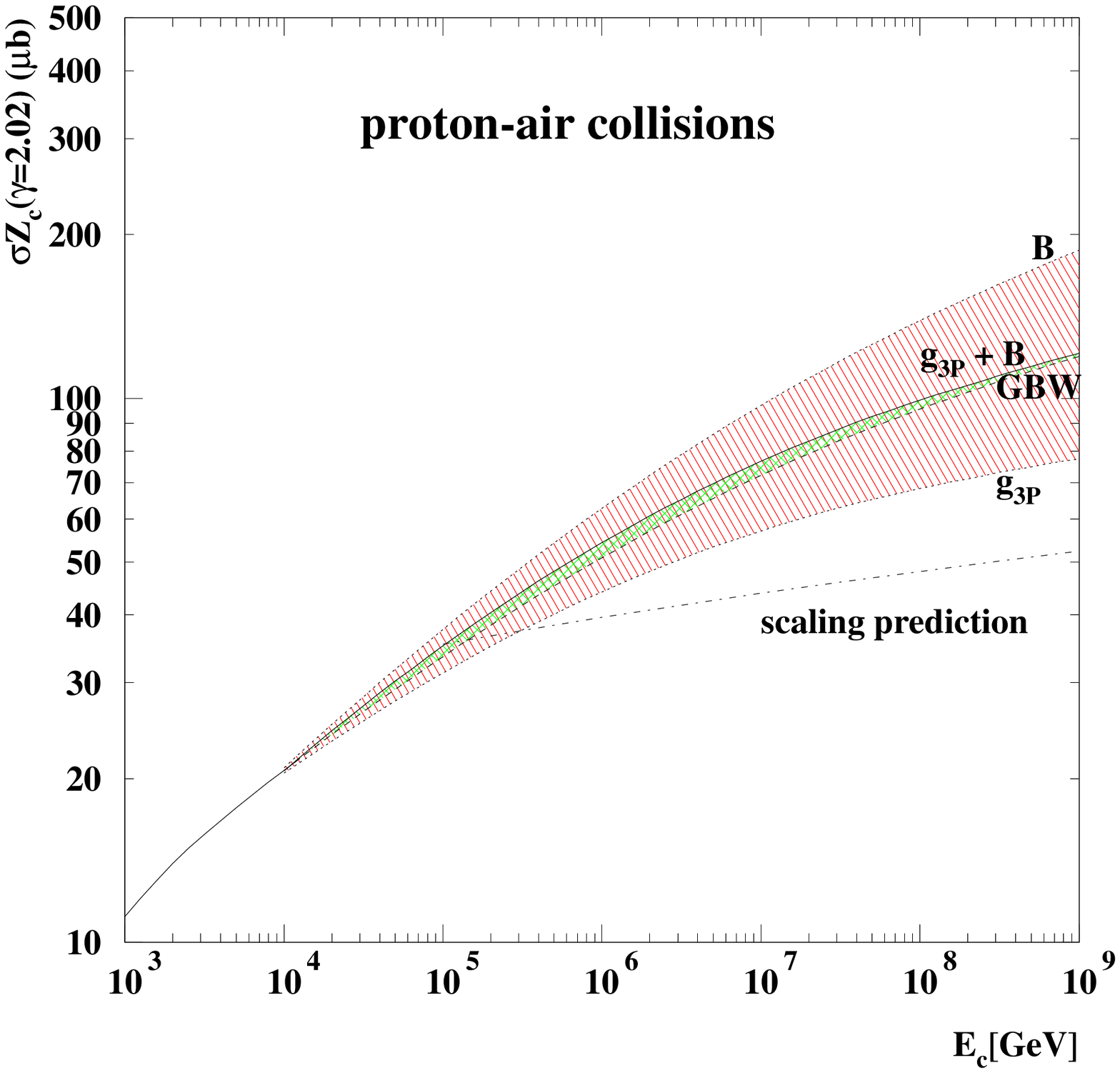,height=5in} \caption{The energy dependence
of the relevant `$Z$ moment', $\sigma Z_c\equiv\int
x^{2.02}(d\sigma^c/dx)dx$, of charm production in $p$-air
collisions, as a function of the energy of the produced charm
quark. The dashed curve corresponds to the GBW model (extended to
include rescattering of the $\cc$ pair within the air nucleus).
The upper dotted ($B$), lower dotted ($g_{3P}$) and continuous
($g_{3P}+B$) curves respectively include the growth of the proton
radius, triple-Pomeron effects and the combination of these two
effects. The dot-dashed curve is the scaling prediction
of~(\ref{eq:lowerlimit}), but with $\sigma_{\rm inel}$
corresponding to $p$--air collisions. The $\gamma=2.02$ moments
are shown for illustration; in the calculation of the neutrino
fluxes, the differential cross section $x_F d\sigma /dx_F$ is
convoluted with the observed primary cosmic ray flux.}
\end{center}\end{figure}
The dashed curve is the prediction of the extrapolation based on
the original GBW saturation model, while the upper dotted curve (marked
by $B$) includes the possible effect of the growth of the proton
radius with energy, as discussed at the end of
Section~\ref{sec:gluonatsmallx}, see in particular (\ref{eq:B}).
Instead of using the Pomeron slope, $\alpha'=0.25\ \GeV^{-2}$,
measured in elastic $pp$ scattering, in (\ref{eq:B}) we use the
value $0.11\ \GeV^{-2}$ together with slope $B=4.4\ \GeV^{-2}$ at
$W=90\ \GeV$, which were deduced from elastic $J/\psi$ production
data at HERA~\cite{Jslope}. These values are appropriate for charm
production. The lower dotted line (marked $g_{3P}$) reflects the
inclusion of an additional absorptive effect---the absorption of
gluons. We take $c=0.2$ in (\ref{eq:R_0^2}), which corresponds to
the largest estimate\footnote{Triple-Pomeron phenomenology gives
$g_{3P}\sim0.5\ \GeV^{-1}$. However, the {\em bare} triple-Pomeron
vertex may be a few times larger, since the phenomenological value
already accounts for some screening effects. It was argued in
Ref.~\cite{BRV} that in perturbative QCD we expect
$g_{3P}=(0.5$--$2)\ \GeV^{-1}$.}, $g_{3P}=2\ \GeV^{-1}$, of the
triple-Pomeron vertex. This choice is made to show the full extent
of the uncertainty in $\cc$ production. If both the above effects
are included (the radius growth and $g_{3P}$), then the
solid curve (marked $B+g_{3P}$) is obtained. Conservatively, we
predict that $Z_c$ will lie within the shaded region in Fig.~4;
the most likely behaviour is that it will follow the cross-hatched
region. Even the conservative prediction has much less uncertainty
than previous estimates. However, for completeness, the dot-dashed
curve shows the extreme lower limit, given by scaling formula
(\ref{eq:lowerlimit}), but where now $\sigma_{\rm inel}$ is the
proton--air cross section. All the variations of the original GBW
model were normalized in the region $E\sim10^5\ \GeV$, where the
partons sampled in the hard subprocess are known, and the model
tuned to the data.

\section{Prompt neutrinos: development of the air shower}
\label{sec:promptneutrinos}

Our aim is to predict the spectra of {\em prompt} $\nu_\mu$ and
$\nu_\tau$ neutrinos produced in the atmosphere by cosmic rays.
Prompt leptons originate from the following sequence: the
production of $\cc$ pairs, their fragmentation into charm hadrons
which then undergo semi-leptonic decay. In the lower energy range
($E<10^7\ \GeV$) it is possible to estimate the leptonic spectra
by simply taking a product of moments of the various
distributions, see, for example, Ref.~\cite{COSTAckbk,CHS}.
However, for $E\gtrsim10^7\ \GeV$ the decay length of $D$ mesons
becomes comparable with the depth of the atmosphere, and so it is
necessary to follow the development of the air shower in more
detail. It is described by a set of equations in terms of the
`depth' $X$ of the atmosphere transversed by a particle in the
vertical direction. $X$ is related to the height $h$ by
\be \label{eq:X} X=\int_h^\infty \rho(h')dh' \ee
where $\rho(h)$ is the density of the atmosphere at vertical
height $h$. We take the same exponential profile of the
atmosphere\footnote{We have checked that the numerical results
essentially do not depend on the precise parameterization of the
density of the atmosphere.} as was used in Ref.~\cite{TIG}. The
sequence of equations determine $\phi_a(E,X)$, which are the
fluxes of the corresponding particles with energy $E$ at depth
$X$, where $a=N,c,i,l$ (that is nucleon, $c$~quark, charmed
hadron, lepton). The initial flux $\phi_N(E,0)$ is the known
primary cosmic ray flux. All other initial fluxes are zero, that
is $\phi_a(E,0)=0$ for $a=c,i,l$. The set of equations which
determine $\phi_N\ra\phi_c\ra\phi_i\ra\phi_l$ are
\be \label{eq:phi_N} \frac{\partial\phi_N(E,X)}{\partial X}\ =\
-\frac{1}{\Lambda_N(E)}\phi_N(E,X); \ee
\be \label{eq:phi_c}\phi_c(E,X)\ =\ \int_E^\infty
dE'dx_c\phi_N(E',X)\frac{n_A}{\rho}\frac{d\sigma}{dx_c}^{p\ra c}
\delta(E-x_cE'); \ee
\be \label{eq:phi_i} \frac{\partial \phi_i(E,X)}{\partial X}\ =\
-\frac{1}{\Lambda_i(E,X)}\phi_i(E,X)+\int_E^\infty
dE'dx\phi_c(E',X)\frac{dn}{dx}^{c\ra i} \delta(E-xE') \ee
with $i=D^\pm,D^0,\bar{D}^0, D_s^\pm, \Lambda_c$; and finally
\be \label{eq:phi_l} \frac{\partial\phi_l(E,X)}{\partial X}\ =\
\sum_i\int_E^\infty dE'dx\phi_i(E',X)\frac{1}{\lambda_i^{\rm
dec}(E',X)}B(i\ra l)\frac{dn}{dx}^{i\ra l} \delta(E-xE') \ee
where $B(i\ra l)$ is the branching fraction of the decay of the
charmed hadron $i$ to lepton $l$. The nucleon attenuation length
is
\be \label{eq:Lambda_N} \Lambda_N(E) \equiv
\frac{\lambda_N(E)}{1-Z_{NN}}\,, \ee
where $Z_{NN}$ is the spectrum-weighted moment for nucleon
regeneration and $\lambda_N$ is the interaction thickness
\be \label{eq:lambda_N}
\lambda_N=\rho(h){\Bigg/}\sum_A\sigma_{NA}(E)n_A(h). \ee
$n_A(h)$ is the number density of air nuclei of atomic number $A$
at height $h$ and $\sigma_{NA}$ is total $NA$ inelastic cross
section. Instead of the sum over $A$ in (\ref{eq:lambda_N}), we
take the mean value $\langle A\rangle = 14.5$ for air. Note that
the factor $n_A/\rho$ in (\ref{eq:phi_c}) arises from
(\ref{eq:lambda_N}). For the nucleon--air cross section,
$\sigma_{N{\rm -}{\rm air}}$, we take the parameterization of
Bugaev et~al.~\cite{BUGAEV}. For the incoming cosmic ray flux
we take the parametrisation given in \cite{COSTAckbk} denoted as TIG with knee.
Also from \cite{COSTAckbk} we take parametrisation of $Z_{NN}$
which depends on energy and takes into account the knee, which is consistent with \cite{TIG}.  This $Z$ factor includes
the regeneration by the $p,n,N^*\dots$ particles.

The attenuation length $\Lambda_i(E,X)$ of the charmed hadrons
consists of two parts: the decay length $\lambda_i^{\rm dec}$ and
attenuation due to their strong interactions with air nuclei. The
decay lengths, $\lambda_i^{\rm dec}$, of the various charmed
hadrons depend, via their $X$ dependence, on the density of the
atmosphere,
\be \label{eq:decaylengths} \lambda_i^{\rm dec} =
c\tau_i\frac{E}{m_i} \rho(X), \ee
where $m_i$ and $\tau_i$, the mass and lifetime of the $i^{\rm
th}$ charmed hadron, are taken from \cite{RPP2002}. The
attenuation due to the strong interactions of the produced charmed
hadrons with the air has a form similar to (\ref{eq:Lambda_N}),
namely $\lambda_i/(1-Z_{\cc})$. To calculate $\lambda_i$ we assume
an absorptive cross section equal to half the absorptive $p$--air
cross section (based on quark counting), and we take a charm
regeneration factor $Z_{\cc}=(0.8)^\gamma$. That is, we estimate
that the leading charm quark carries a fraction
$x=m_c/(m_c+m_q)\simeq 0.8$ of the incoming energy, where
$m_q\simeq0.3\ \GeV$ is the mass of a light constituent quark. For
simplicity, we take the same $Z_{\cc}$ and $\lambda_i$ for all
charm hadrons; for $\Lambda_c$ we expect the larger cross section
to be approximately compensated by the larger $Z_{\cc}$. Thus,
finally, $\Lambda_i$ is given by
\be \label{eq:Lambda_i} \frac{1}{\Lambda_i}\ =\
\frac{1}{\lambda_i^{\rm dec}} + \frac{1-Z_{\cc}}{\lambda_i}. \ee

From (\ref{eq:phi_N}) the light baryon flux is given by
\be \label{eq:lightbaryonflux} \phi_N(E,X) =
\phi_N(E,0)\exp(-X/\Lambda_N(E)). \ee
If we insert (\ref{eq:phi_c}) into (\ref{eq:phi_i}), then the
individual charm hadron fluxes are
\be \label{eq:hadronfluxes} \phi_i(E,X) = \int_0^X
dX'\exp\left(-\int_{X'}^X \frac{dX''}{\Lambda_i(E,X'')}\right)
S_{N\ra i}(E,X') \ee
where
\be \label{eq:S_Ntoi} S_{N\ra i}(E,X) = \int_E^\infty dE'
\frac{n_A}{\rho} \frac{1}{E'}\frac{d\sigma}{dx}^{N\ra
i}\raisebox{1.58ex}{$(E,E')$}\phi_N(E',X) \ee
and $x=E/E'$.

The spectra of charmed hadrons $d\sigma^{N\ra i}/dx$ are
calculated using the three models shown in Fig.~2. The ratios of
the different charm yields (after the hadronization of the $c$
quark) are given in Ref.~\cite{FRIX} to be
\be \label{eq:charmyields}
\frac{\sigma(D_s^+)}{\sigma(D^+,D^0)}=0.2, \quad
\frac{\sigma(\Lambda_c)}{\sigma(D^+,D^0)}=0.3,\quad
\frac{\sigma(D^+)}{\sigma(D^0)}=0.5. \ee
More recent data~\cite{DsDATA} favour a smaller value\footnote{For
isolated $c$-quark fragmentation, PYTHIA~\cite{PYTHIA} gives a
ratio of 0.08.} of the first ratio quoted in
(\ref{eq:charmyields}), namely
\be \label{eq:smallersigmaratio}
\frac{\sigma(D_s^+)}{\sigma(D^+,D^0)} \simeq 0.1. \ee
We take this value in our analysis.

Note that the $\Lambda_c$ baryon is produced by the recombination of a $c$ quark with a spectator diquark of the
incoming nucleon\footnote{The PYTHIA/JETSET Monte Carlo gives only 2.5\% $\Lambda_c$ baryons in the fragmentation
of an isolated $c$ jet, that is in the absence of spectator diquarks.}. It is not produced from a $\bar{c}$ quark.
Therefore the parton momentum fraction $x_\Lambda$ carried by the $\Lambda_c$ is $x_\Lambda = x_d + x_c$. The
diquark momentum fraction $x_d$ will be less than $\frac{2}{3}(1-x_c)$, as part of the energy is carried away by
the third valence quark (the factor $\frac{2}{3}$), by the $\bar c$ quark and by gluons. To allow  for this, we
therefore take $x_d = \frac{1}{2}(1-x_c)$, which leads to
\be \label{eq:prescriptionA} x_\Lambda =  {\textstyle \frac{1}{2}}(1+x_c) \ee
This is found to be in good agreement with the distribution generated by the PYTHIA Monte Carlo~\cite{PYTHIA},
which has a maximum in the region $x_\Lambda \simeq 0.5$--$0.6$. Of course a very slow $c$ quark is unlikely to
combine with a fast light diquark (with $x_d \sim 0.5$). Therefore we introduce a cut-off, $x_c>x_c^0$,
in~(\ref{eq:prescriptionA}). Assuming a mean velocity of the $c$ quark to be $\langle\,v^2\,\rangle\sim0.25$, we
estimate $x_c^0\sim0.1$. For $\Lambda_b$ production, which we discuss below, the heavier $b$ quark will carry a
larger fraction of the $\Lambda_b$ momentum. In this case we take the cut-off to be $x_b^0 = 0.25$. As a check, we
also compute the prompt flux, arising from $c\to\Lambda_c$, using an alternative to
prescription~(\ref{eq:prescriptionA}). We assume that for the diquark $x_d= (m_d/m_c)/x_c$ with a constituent
diquark of mass $m_d = 2m_q = 0.7$~GeV and $m_c = 1.5$~GeV. With this assumption
\be \label{eq:prescriptionB} x_\Lambda = 1.47x_c, \ee
or, in the case of beauty, $x_\Lambda = 1.16 x_b$, using $m_b = 4.5$~GeV.


Rather than using a fragmentation function for $dn^{c\ra D}/dx$, for $D$ mesons we take $x_D=0.75 x_c$. This is
sufficient for our purposes\footnote{That is the distributions $dn^{c\ra i}/dx$ were taken to be proportional to
$\delta(x_D-0.75x_c)$ for $D$ mesons, whereas for $\Lambda_c$ we assume that they are proportional to
$\delta(x_\Lambda-\frac{1}{2}(x_c+1))$ for $x_c>0.1$, or $\delta(x_\Lambda - 1.47x_c)$. We note that PYTHIA gives
a harder $x=E_D/E_c$ distribution than that shown for the Peterson et~al. function~\cite{PET} in
Ref.~\cite{RPP2002}.}. For illustration we compare, in Fig.~5, the prompt $\nu_\mu + \bar\nu_\mu$ flux at ground
level, $\phi_{\nu_\mu+\bar\nu_\mu}(E)$, obtained from~(\ref{eq:phi_l}) using different forms of fragmentation.
Clearly the upper curve, corresponding to no fragmentation, gives an overestimate of the flux.  Moreover, due to
the presence of additional light ``sea'' quarks, we expect a harder distribution for the fragmentation in $pp$
production than that obtained in $e^+e^-$ collisions (lower curve). Hence our use of fragmentation corresponding
to one of the middle curves. We see that both the $\Lambda_c$ hadronization prescriptions~(\ref{eq:prescriptionA})
and (\ref{eq:prescriptionB}) give very similar fluxes. At the highest energies the fraction of neutrinos coming
from $\Lambda_c$ (relative to those from $D$) increases due to the short $\Lambda_c$ lifetime. Therefore the
choice $x_\Lambda = \frac{1}{2}(1+x_c)$ of (\ref{eq:prescriptionA}), which corresponds to a larger
$\langle\,x_\Lambda\,\rangle$ than (\ref{eq:prescriptionB}), gives a larger neutrino flux for
$E_\nu\gtrsim10^6$~GeV. The results below correspond to using prescription~(\ref{eq:prescriptionA}).

For the leptonic decay of each charm hadron $i$, the distribution
$dn^{i\ra l}/dx_l$ was generated by the PYTHIA Monte
Carlo~\cite{PYTHIA}. Note that in PYTHIA~6.2 the
$D_s\ra\tau\nu_\tau$ branching ratio was set to be 1\%, whereas
the latest value is $6.4\pm1.5\%$~\cite{RPP2002}. Since $D_s\ra
\tau\nu_\tau$ is almost the only source of prompt $\nu_\tau$
neutrinos, it is important to renormalise the yield using the
updated branching ratio.  It is interesting to note that the
$D_s^+$ decay produces more prompt $\bar{\nu}_\tau$ neutrinos than
$\nu_\tau$, since the $\bar{\nu}_\tau$ from $\tau^+$ decay has the
large $x_l$. Of course, the reverse is true for $D_s^-$ decay.

\section{Prompt $\nu_\mu$ and $\nu_\tau$ fluxes} \label{sec:promptnufluxes}

We present the predicted yields of prompt $\nu_\mu$ and $\nu_\tau$
neutrinos from charmed hadrons produced by cosmic ray collisions
in the atmosphere. The flux of $\nu_e$ neutrinos is essentially
equal to that of $\nu_\mu$ neutrinos. High energy prompt electrons
are completely degraded in the electromagnetic cascade. For prompt
muons the electromagnetic interaction is much weaker; it was
demonstrated in Ref.~\cite{GGV4} that the prompt $\mu$ flux is
only about 10\% smaller than the prompt $\nu_\mu$ flux at the
surface of the earth.

\begin{figure}[h]
\begin{center}
\epsfig{figure=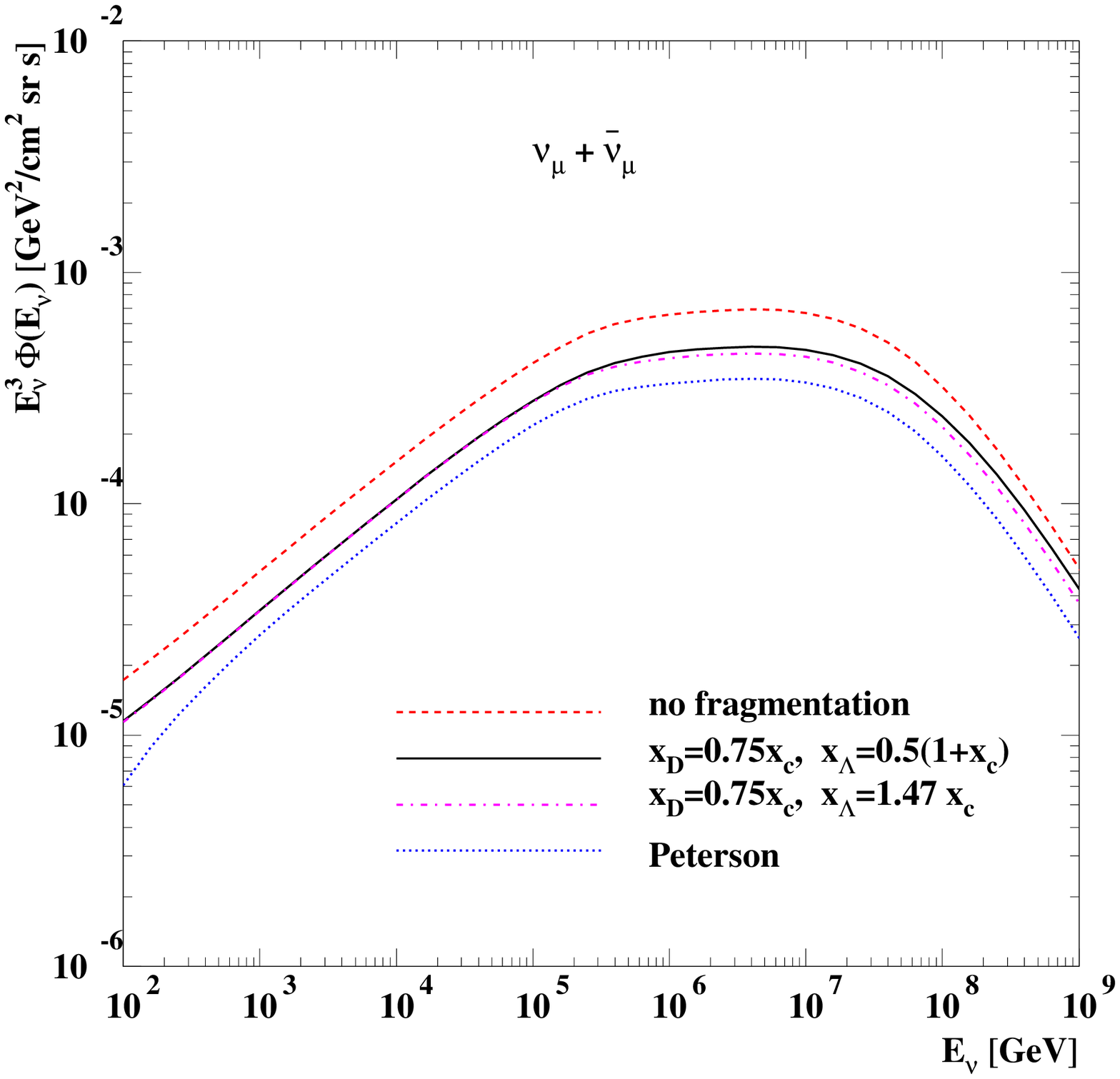,height=5in} \caption{The flux of prompt muon neutrinos at ground level, weighted by $E^3$,
for different choices of the $c\ra$\ charmed hadron fragmentation function, $dn^{c\ra i}/dx$. The curves
correspond in descending order to assuming (i)~no fragmentation $dn/dx\propto\delta(1-x)$,
(ii,$\,$iii)~$\delta(x_D-0.75x_c)$ and $\delta(x_{\Lambda_c} - \frac{1}{2}(x_c+1))$ for $x_c>0.1$, or
$\delta(x_{\Lambda_c} - 1.47x_c)$ and (iv)~a Peterson et~al. fragmentation function~\cite{PET} with $\varepsilon_c
= 0.043$~\cite{RPP2002}. In each case charm production is calculated using the GBW solid curve of Fig.~4.}
\end{center}\end{figure}
\begin{figure}[h]
\begin{center}
\epsfig{figure=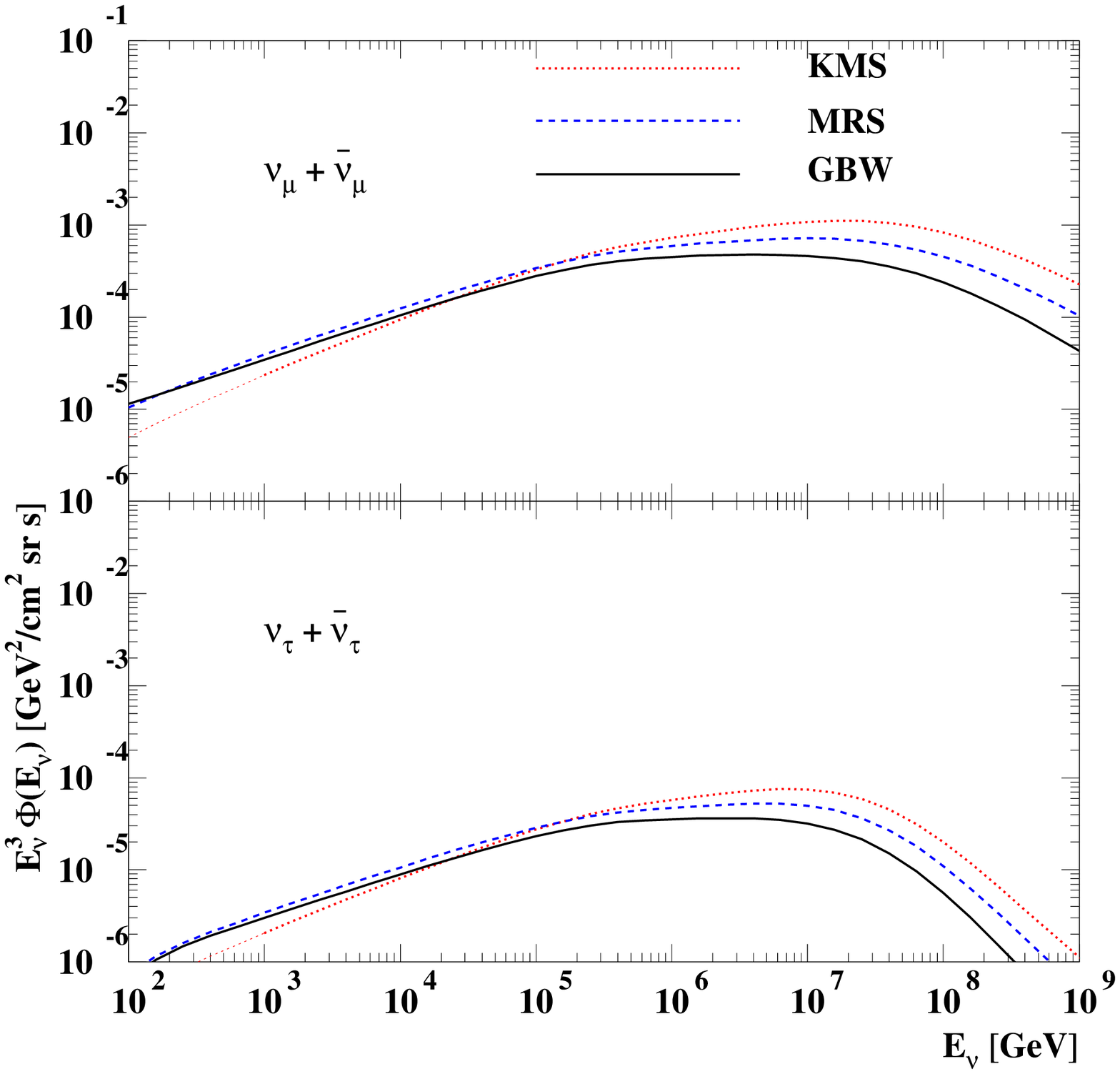,height=5in} \caption{The prompt
$\nu_\mu+\bar{\nu}_\mu$ and $\nu_\tau+\bar{\nu}_\tau$ fluxes,
arising from $\cc$ production, fragmentation and decay, obtained
using the three different extrapolations of the gluon to very
small $x$ (described in the caption to Fig.~2). For the MRST and
KMS models, the $A$ dependence is taken to be $d\sigma/dx(p+{\rm
air}\ra c+\dots) = Ad\sigma/dx(pN\ra c+\dots)$.}
\end{center}\end{figure}

In Fig.~6 we plot the prompt $\nu_\mu$ and $\nu_\tau$ fluxes
predicted by the three models for the extrapolation of the gluon
distribution to very small $x$. Although in Figs.~3 and 4 we
compared different models of extrapolation using a fixed
$\gamma=2.02$, in the actual calculation of the neutrino fluxes we
used the observed primary cosmic ray flux, which corresponds to
different values of $\gamma$ above and below the `knee'.  The
sharp fall-off for neutrino energies $E>10^8~\GeV$ is due to the
increase in the decay length of the charmed hadrons, arising from
Lorentz time dilation. Clearly in the gluon kinematic domain of
interest ($x<10^{-5}$ and $Q^2\lesim10\ \GeV^2$) saturation
effects become important. The reason for the behaviour of the KMS
prediction---below at low energies, above at high energies---was
given in the discussion concerning Fig.~3. We have argued that
extrapolations based on the GBW model and its variations, as shown
in Fig.~4, give the most reliable predictions for $E>10^6\ \GeV$.
Thus in Fig.~7 we show the spread of predictions of the neutrino
fluxes based on the shaded domain in Fig.~4. In Fig.~7 we also
show the conventional atmospheric flux of $\nu_\mu$ (from $\pi$
and $K$ decays, etc.). Moreover, there is a small probability that
atmospheric $\nu_\mu$ neutrinos may oscillate into $\nu_\tau$
neutrinos and so provide an `atmospheric' $\nu_\tau$ flux. We
calculate this flux using the $3\sigma$ ranges of the
$\sin^2\theta_{\rm ATM}$ and $\Delta m^2_{\rm ATM}$ neutrino
mixing parameters found in an analysis~\cite{VALLE} of the
Super-Kamiokande~\cite{SK} and MACRO~\cite{MACRO} data. The
resulting atmospheric $\nu_\tau$ flux is shown by a band in the
lower plot of Fig.~7.
\begin{figure}[h]
\begin{center}
\epsfig{figure=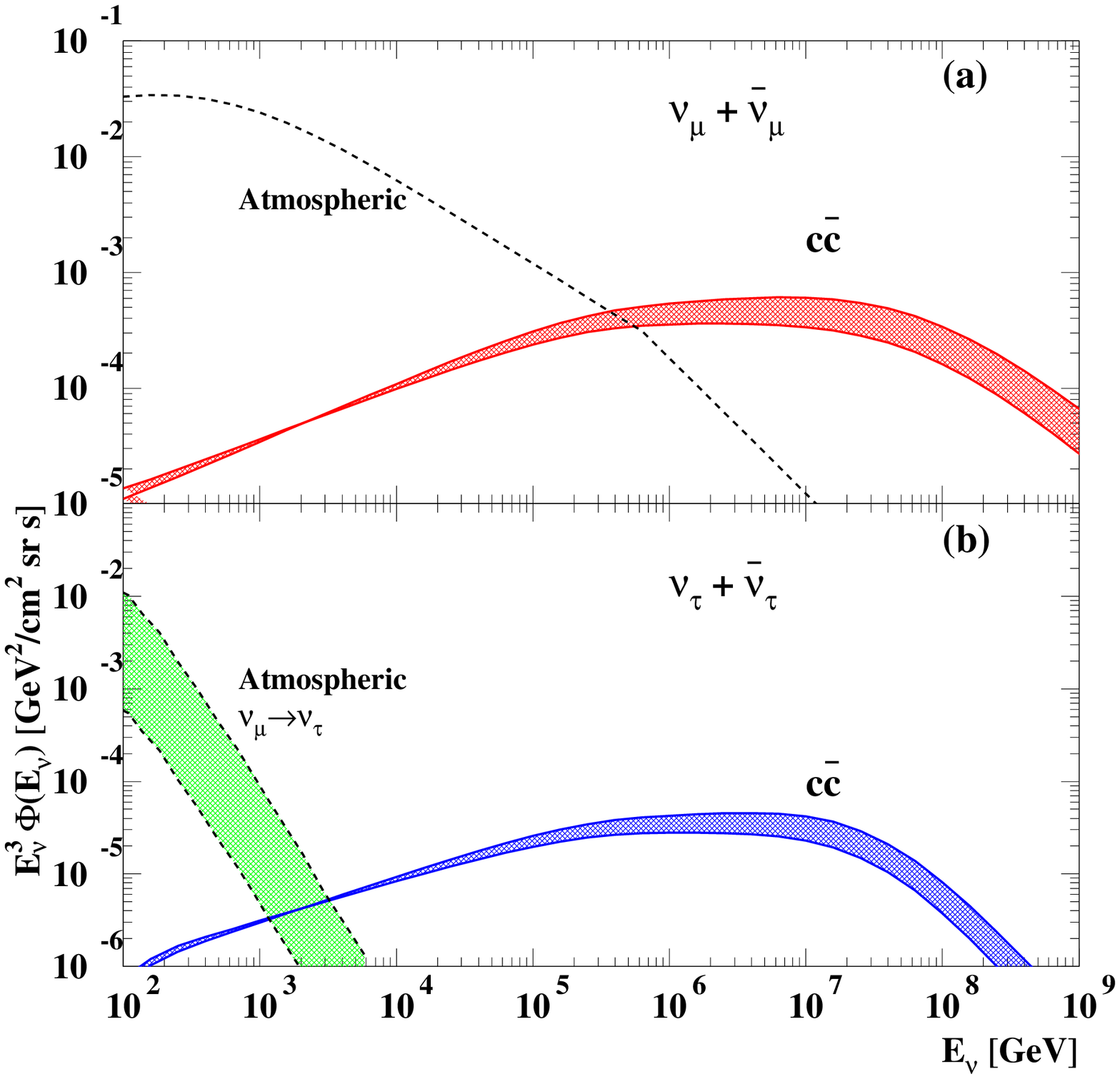,height=5in} \caption{The prompt
(a)~$\nu_\mu+\bar{\nu}_\mu$ and (b)~$\nu_\tau+\bar{\nu}_\tau$
fluxes calculated using the charm production cross sections
corresponding to the shaded band in Fig.~4. Also shown are the
conventional muon and tau neutrino atmospheric fluxes, where the
latter originates, via neutrino mixing transitions, from the
former. There is also a contribution to the prompt $\nu_\tau +
\bar{\nu}_\tau$ flux from beauty production, which is not included
here, but is shown in Fig.~10(b). The prompt $\nu_e+\bar\nu_e$
flux is equal to that for $\nu_\mu + \bar\nu_\mu$, but the
atmospheric flux is a factor of 10 or so less, see the discussion
in Section~8.1.}
\end{center}\end{figure}

We discuss the neutrino fluxes of Fig.~7 in the concluding
section, after we have included the contribution to the prompt
$\nu_\tau$ spectrum arising from $\bb$ production, fragmentation
and decay. However, first, we show in Fig.~8 the effect of charmed
hadron interactions with the atmosphere.
\begin{figure}[h]
\begin{center}
\epsfig{figure=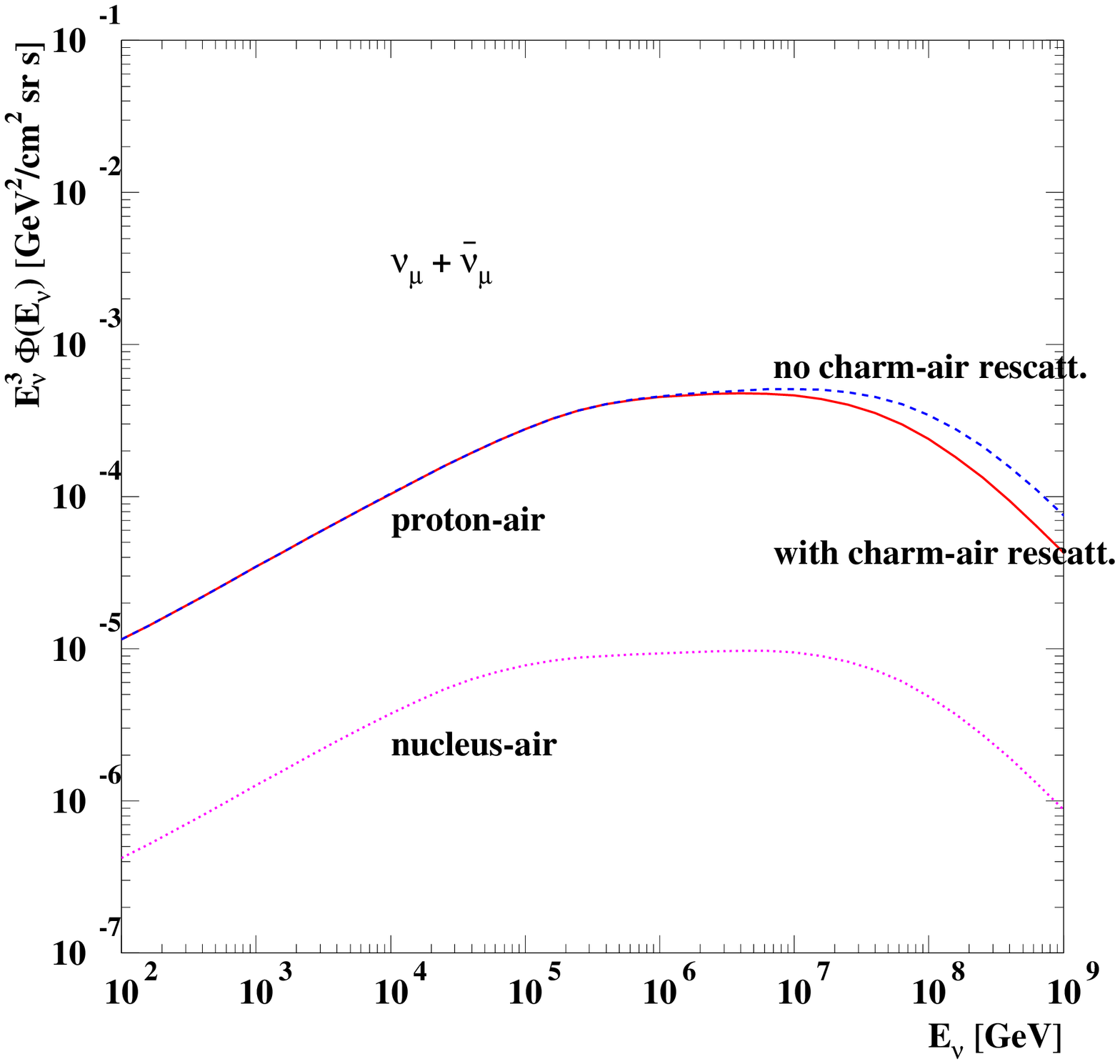,height=5in} \caption{The dashed and dotted curves correspond, respectively, to the prompt
$\nu_\mu+\bar{\nu}_\mu$ flux obtained by switching off the charmed hadron--air interactions and assuming that the
incoming cosmic rays have $\langle A \rangle = 7$ rather than $A=1$. The default continuous curve (with air
interactions and $A=1$) is based on the GBW gluon extrapolation. In this work we assume that the $c\to\Lambda_c$
hadronization is given by $x_\Lambda = \frac{1}{2}(1+x_c)$ for $x_c>0.1$.}
\end{center}\end{figure}
The effect is illustrated by the difference
between the dashed curve, for which the interactions are
suppressed (that is $\lambda_i=0$), and the continuous curve with
the interactions present. For this comparison we use the GBW
extrapolation, the one given by the solid curve in Fig.~4.

We emphasize that the predictions for the prompt neutrino fluxes
depend strongly on the nuclear content of the primary cosmic rays.
So far we have assumed that the cosmic rays are composed entirely
of protons. Suppose that the protons were to be replaced by nuclei
of the same energy $E$ and of atomic number $A$. Then we have to
scale the energy of the primary interacting nucleon to $E/A$.
Roughly speaking, this reduces the neutrino flux $\phi_\nu$ in the
plateau region ($10^6$--$10^7\ \GeV$) of the $E^3\phi_\nu$ plot by
$A^3$. On the other hand, the number of incoming nucleons is $A$,
so as a consequence we expect an $A^2$ suppression. A detailed
calculation for incoming nuclei with $\langle A \rangle =
7$~\cite{CHBS} gives the dotted curve in Fig.~8. The suppression
of the original GBW continuous curve (for $A=1$) is apparent.

\section{Prompt $\nu_\tau$ flux from $\bb$ production}\label{sec:prompt}

At first sight it appears that if we also allow for $\bb$
production, with the same cross section, fragmentation and decay
then we will approximately double the $\nu_\tau$ flux. However
note that, in comparison with charm production, the cross section
for beauty production is about 30 times smaller due to the factor
$m_c^2/m_b^2$ and to the larger value of $x_2$ of the gluon
structure function which is sampled, see
eq.~(\ref{eq:dsigmabydx_F}), see Fig.~9. Nevertheless, the high
energy production of $\nu_\tau$ neutrinos from beauty decays is
not negligible.
\begin{figure}[h]
\begin{center}
\epsfig{figure=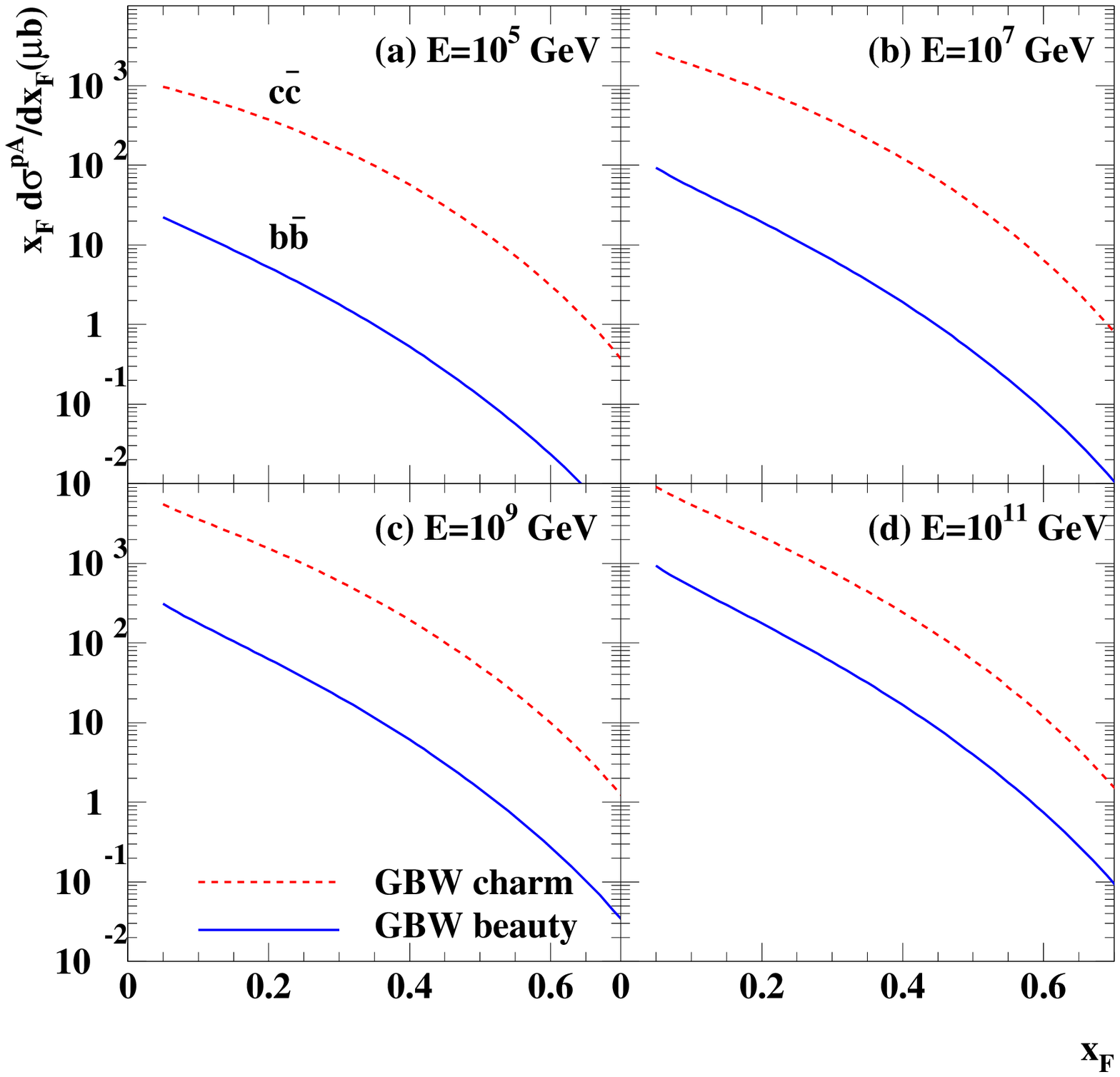,height=5in} \caption{The differential
cross section $x_F\,d\sigma/dx_F$ for charm and beauty production
in p--air collisions for four different laboratory energies $E$.
For $\bb$ production we take $m_b=4.5$~GeV.}
\end{center}\end{figure}
For charm, only the decay of the $D_s$ may produce $\nu_\tau$
neutrinos, whereas now $B^\pm$, $B^0$, $B_s$ and $\Lambda_b$
semileptonic decays also give rise to a significant $\nu_\tau$
flux. Indeed including $B$ and $\Lambda_b$ decays  enlarges the
predicted total prompt $\nu_\tau$ flux by about 40\% for
$E_\nu\sim10^5~\GeV$, and even more at higher energies, see
Fig.~10(b). This is considerably larger than the $\nu_{\tau}$ flux
calculated in \cite{PRTAU}, where the beauty induced contribution
grows from 1\% at $E_\nu = 10^2$~GeV to about 10\% at
$E_\nu=10^6$~GeV. The reasons why we obtain a larger fraction of
$\nu_\tau$ neutrinos from beauty are, first, that we use
(\ref{eq:smallersigmaratio}), rather than (\ref{eq:charmyields}),
for $c\ra D_s$ hadronization and, second, our cross section for
charm production is more suppressed at high energies by absorptive
corrections, than the more compact $\bb$ production process.

Fig.~10(a) shows the breakdown of the total prompt $\nu_\tau$
flux. We take the ratios of the $B^\pm$, $B^0$, $B_s$ and
$\Lambda_b$ beauty yields (after hadronization of the $b$ quark)
to be given by exactly analogous relations to
(\ref{eq:charmyields}) and (\ref{eq:smallersigmaratio}) for charm.
Recall that $\nu_\tau$'s of $\cc$ and $\bb$ origin come,
respectively, from $D_s\ra\tau\nu_\tau$ and from the semileptonic
$\tau\nu_\tau$ decay modes of $B^\pm$, $B^0$, $B_s$ and
$\Lambda_b$. Thus we have a direct $\nu_\tau$ component and an
indirect component coming from the $\tau\ra\nu_\tau$ decays. For
charm the former component is very small since the direct
$\nu_\tau$ carries away a small energy fraction. Thus the
$\tau\ra\nu_\tau$ decay gives the dominant component, which is
truncated at an energy when the $\tau$ has insufficient time to
decay. For beauty the direct and indirect components are
comparable until the energy regime is entered at which the $\tau$
has not enough time to decay.

To calculate the $b$ quark cross section we have used the same GBW
model as described in Sections~2--4 with a quark mass
$m_b=4.5~\GeV$. We have approximated the $b\ra B$ fragmentation
function by a delta function at $x=E_B/E_b=0.9$. Also we have
assumed that the hadronization and the cross-section of the beauty
hadron--air interaction do not depend appreciably on the nature of
the heavy quark. So we have used the same attenuation lengths
$\Lambda_i$ and the same ratios of the different beauty hadron
yields (eqs.~(\ref{eq:charmyields}) and
(\ref{eq:smallersigmaratio})), as were used for the charm case.

We neglect the $\bb$ contribution to the $\nu_\mu$ (and $\nu_e$)
fluxes. They never exceed about 2--3\% of that arising from $\cc$
production and decay.

\section{Concluding discussion} \label{sec:concluding}

\subsection{Implications for Physics} \label{sec:implications}

The $\nu_\mu$ flux shown in Fig.~7(a), and the $\nu_\tau$ flux
shown in Fig.~10(b), have important implications for neutrino
astronomy. These neutrino fluxes arise from cosmic ray
interactions with the atmosphere. They therefore provide the
background to searches for cosmic neutrinos (for which there are
many exciting New Physics scenarios). They also provide a
possibility to calibrate the detectors of the new neutrino
telescopes. A particularly interesting energy domain is where the
prompt neutrino flux has emerged from the sharply falling
`conventional' atmospheric neutrino flux. Here $\nu_\tau$ appears
to have an enormous advantage in searching for a signal for New
Physics, see Figs.~7 and~10. Due to mixing, the fluxes of
$\nu_\mu$ and $\nu_\tau$ are equal for incoming cosmic neutrinos.
The flux of prompt $\nu_\tau$ neutrinos (which arises mainly from
$D_s\ra\tau\nu_\tau$ decays, but with a significant 40\%
contribution from $B$, $B_s$, $\Lambda_b$ semileptonic decays) is
about ten times less than that for prompt $\nu_\mu$, and the
number of $\nu_\tau$ neutrinos produced from the conventional
atmospheric flux (via $\nu_\mu \ra \nu_\tau$ oscillations) is
negligible for $E>10^4\ \GeV$. There is only a little time for
neutrino mixing in the atmosphere. Thus the tau neutrino flux
originating from atmospheric neutrinos is greatly suppressed. As a
consequence, tau neutrinos offer an ideal means of identifying
neutrinos of {\em cosmic} origin, and for searching for New
Physics.

There is another reason to concentrate on the $\nu_\tau$ flux. For
values of $E_\nu$ above $10^4$~GeV the $\nu_\mu$ flux is
significantly depleted in the passage of the neutrinos through the
Earth. The absorption increases rapidly as $E_\nu$ is
increased~\cite{NUABS}. On the other hand, high energy $\nu_\tau$
neutrinos have the unique advantage that they are not depleted in
number no matter how much of the Earth that they pass through. If
they suffer a charged-current interaction, they produce a $\tau$
lepton which subsequently decays, regenerating a $\nu_\tau$
neutrino with degraded energy~\cite{NUTAU}. It is, moreover,
interesting to note that tau neutrinos of energies about
$10^7$~GeV may produce spectacular distinctive signatures in the
planned 1~km$^3$ high energy neutrino detectors made of strings of
photomultiplier tubes distributed throughout a naturally occurring
Cerenkov medium, such as water or ice, deep in the ocean or ice
cap. A charged-current $\nu_\tau$ interaction may produce a
contained `double bang' signature~\cite{DBANG} or a `lollipop'
signature~\cite{LOLLI}. In the former, the first bang corresponds
to the hadronic shower produced along with the subsequent $\tau$
lepton and the second bang is the shower associated with the
$\tau$ decay. A lollipop  event is when only the second shower
occurs within the detector and the $\tau$ lepton which initiates
the shower is identified by the relatively weak ionization that it
causes.

Another interesting possibility follows from the observation that electron neutrinos may be distinguished
experimentally from muon and tau neutrinos. Electron neutrinos give rise to distinctive showers, which result from
their charged-current interactions, which characteristically are contained within the large detector volume, and
hence identified. The conventional atmospheric flux of $\nu_e$ neutrinos comes mostly from kaon decays and so the
relative flux $\nu_e/\nu_\mu\lesim 0.1$~\cite{GHON}. On the other hand, the prompt $\nu_e$ and $\nu_\mu$ fluxes
are practically equal to each other. Hence the prompt $\nu_e$ flux should be more visible over its conventional
background, than the prompt $\nu_\tau$ flux\footnote{We thank John Beacom for drawing this possibility to our
attention.}. Of course, the background reduction is much less than for $\nu_\tau$ neutrinos, but this will be
partially offset since it is likely that $\nu_e$ neutrinos will be easier to identify than $\nu_\tau$ neutrinos.

So far we have shown the prompt neutrino fluxes arising from
vertically-incident cosmic rays. However, the predicted fluxes are
sensitive to the zenith angle. In particular, as we depart from
the vertical incident direction, more atmosphere is encountered,
which gives more time for the charmed hadrons to decay and hence
allows more high energy neutrinos to be produced. This is well
illustrated by Fig.~11, which shows the prompt neutrino fluxes
produced by horizontally-incident cosmic rays.
\begin{figure}[h]
\begin{center}
\epsfig{figure=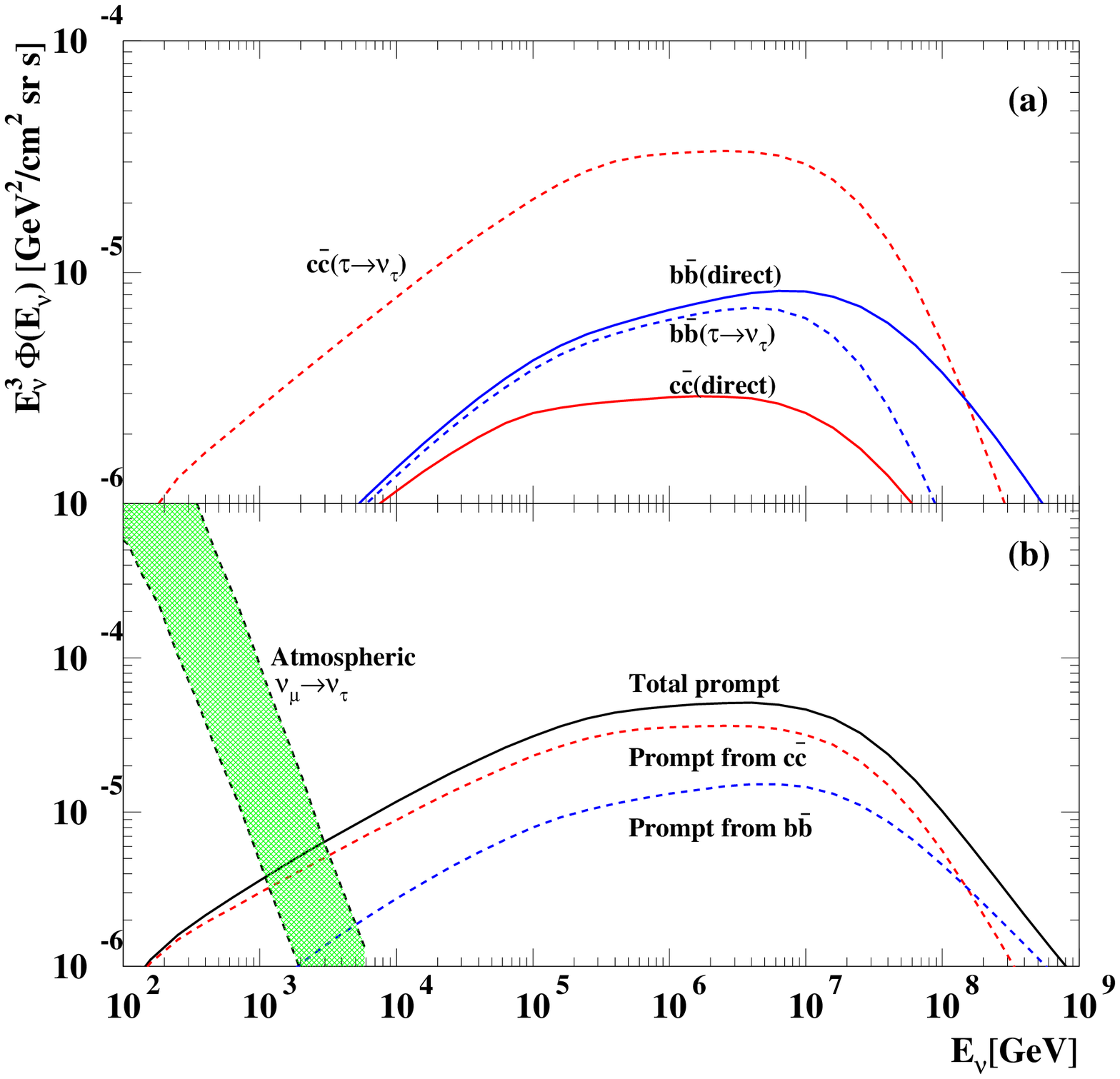,height=5in} \caption{The prompt
$\nu_\tau+\bar \nu_\tau$ fluxes originating from $\cc$ and $\bb$
production and decay, which respectively arise from the
$D_s\ra\tau\nu_\tau$ decay, and from the $B^\pm$, $B^0$, $B_s$ and
$\Lambda_b$ semileptonic $\tau\nu_\tau$ decay modes. The upper
plot shows the breakdown into the direct $\nu_\tau$ contribution
(continuous curves) and the indirect $\tau\ra\nu_\tau$
contribution (dashed curves). The lower plot shows the total
prompt $\nu_\tau + \bar{\nu}_\tau$ flux, together with its
components of $\cc$ and $\bb$ origin. Also shown is the non-prompt
$\nu_\tau + \bar{\nu}_\tau$ flux arising from $\nu_\mu \ra
\nu_\tau$ oscillations from the conventional atmospheric $\nu_\mu$
flux.}
\end{center}\end{figure}
\begin{figure}[h]
\begin{center}
\epsfig{figure=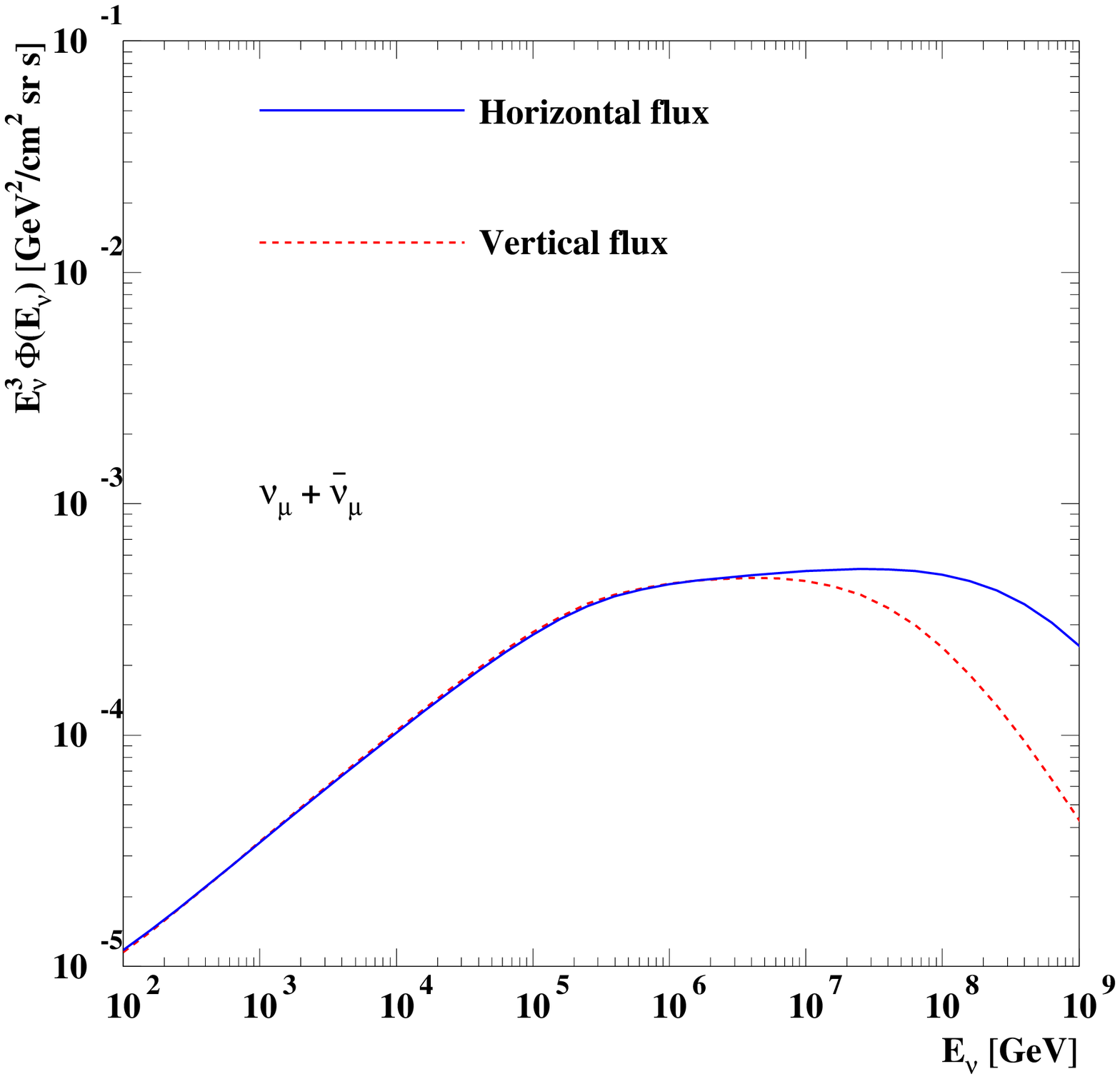,height=5in} \caption{The continuous curve
is the prompt $\nu_\mu+\bar{\nu}_\mu$ flux from
horizontally-incident cosmic rays. The dashed curve, which
corresponds to vertically-incident cosmic rays, is taken from
Fig.~7.}
\end{center}\end{figure}
Indeed, for $E_\nu>10^8~\GeV$, the flux in the horizontal
direction is noticeably larger. The observation of the zenith
angle dependence will allow the `atmospheric' background neutrinos
to be identified and hence cosmic neutrinos (and therefore New
Physics) to be isolated.

\subsection{Uncertainties in the predicted prompt $\nu_\mu$ and $\nu_\tau$ fluxes}

We have emphasized the importance to neutrino astronomy of
reliable predictions of the prompt muon and tau neutrino fluxes,
which arise from $\cc$ and $\bb$ production, hadronization and
decay in the atmosphere. They provide the background to the search
for cosmic neutrinos originating from Active Galactic Nuclei or
elsewhere. We have argued that the predictions based on
perturbative QCD, given in Figs.~7(a) and 10(b), have much less
uncertainty than those that already exist in the literature. How
reliable is perturbative QCD and the extrapolation of the gluon
into the very small $x$ domain? Perturbative QCD is expected to be
valid for $\bb$ production\footnote{Unlike $\cc$ production, there
are no accelerator $\bb$ data in the fragmentation region which is
relevant to this study. However, there exist Tevatron data for
central $\bb$ production with, typically,
$p_{bT}>5$~GeV~\cite{BBBAR} which appear to exceed the NLO QCD
prediction. Recently it has been shown~\cite{NASON} that the
discrepancy may be reduced to an acceptable level by using a
fragmentation function with $\langle x \rangle$ larger than that
of the conventional Peterson et~al. parameterization~\cite{PET}.},
and should also hold for $\cc$ production since the factorization
scale $\mu_F\sim m_c$. It is important to note that at fixed
target energies, $E\sim250$~GeV, the NLO predictions for
$d\sigma/dx_F$ are in agreement with the available data, while for
very high energies the mean scale, $\langle 1/r \rangle$,
increases due to saturation effects. Since the gluon distribution
is determined by HERA deep inelastic data down to $x\sim 10^{-4}$,
predictions, which use partonic structure determined from these
data and which agree with fixed target charm data, should be
reliable up to about $E\sim10^5$~GeV, and perhaps an order of
magnitude or so above. Fig.~3 shows examples of three very
different models which illustrate that this is indeed the case.
(Recall that in Section~\ref{sec:predictions} we explained why the
KMS prediction is not applicable at the lower energies shown in
Fig.~3.) However, as we proceed to higher energies, and sample
smaller and smaller $x$ values and hence increasing gluon density,
we must include the effects of saturation. Though we have labelled
the predictions by GBW, we have not simply used the saturation
model of Golec-Biernat and W\"usthoff, but rather we investigated
the uncertainties in including saturation effects in some detail,
see Fig.~4. We concluded that the most reliable predictions would
be obtained using the continuous curve in Fig.~4, with
conservative errors shown by the shaded band, which are reflected
in the shaded bands on the prompt neutrino fluxes shown in Fig.~7.
These bands represent the uncertainty in $\cc$ production.

In addition, there are also uncertainties associated with the fragmentation of the charm quark (see Fig.~5) and
with the $D$ meson attenuation due to its strong interaction with the atmosphere (see Fig.~8). Another source of
uncertainty is the $x$ distribution of the $\Lambda_c$ hyperons. The recombination of the $c$ quark with the $ud$
diquark of the incoming proton gives the $\Lambda_c$ a rather large $x$. We use the simplified formula $x_\Lambda
= \frac{1}{2}(1+x_c)$, where a fast diquark with $x_{ud}\sim0.5$ recombines with any $c$ quark with $x_c>0.1$. The
relatively large uncertainty arises because the $\nu$ flux is proportional to the $x^{\gamma=2.02}$ moment,
(\ref{eq:momentZ_c}), of the $x$ distribution. Thus the larger value of $x$ carried by the $\Lambda_c$ (as
compared to the $D$ mesons) compensates for the small probability of $\Lambda_c$ formation, see
(\ref{eq:charmyields}).   As a consequence, the contribution of the $\Lambda_c$  to the prompt $\nu_\mu$ flux
increases from about 10\% to 40\% as the neutrino energy increase from $10^4$ to $10^9$~GeV.
 Unfortunately the $x$ distribution of $\Lambda_c$ hyperons at high energies ($E>10$~TeV)
is not measured yet. However, the $x^{2.02}$ moment given by PYTHIA is in agreement with the above prescription,
(\ref{eq:prescriptionA}), that $x_\Lambda = \frac{1}{2}(1+x_c)$ for $x_c>0.1$.

Overall, we see that the uncertainty in the prompt neutrino flux predictions is about a factor of~3. For the
$\nu_\tau$ flux from charm, which relies entirely on the $D_s\ra\tau\nu_\tau$ decay, there is also an extra
uncertainty associated with the branching ratio and with the production ratio of (\ref{eq:smallersigmaratio}).
There is less uncertainty in the $\nu_\tau$ flux from beauty, since it originates more uniformly from $B^\pm$,
$B^0$, $B_s$ and $\Lambda_b$ semileptonic $\tau\nu_\tau$ decays.

Finally we note that the predictions have been made assuming that
the incoming cosmic rays are predominantly composed of protons.
If, however, it should happen that the observed $\nu_\mu$ and
$\nu_\tau$ prompt fluxes lie below the predictions of Figs.~7(a)
and 10(b), then the measurements may give important constraints on
the nuclear composition of cosmic rays.

\section*{Acknowledgements}

It is a special pleasure to acknowledge that over many years we
have benefited from many valuable and illuminating discussions
with Jan Kwieci\'nski, on this and many other topics. We thank
Francis Halzen for drawing the problem of prompt neutrinos to our
attention. A.M.S. thanks Leonard Le\'sniak for discussions. This
work was partially supported by the UK Particle Physics and
Astronomy Research Council, the Russian Fund for Fundamental
Research (grant 01-02-17095), the British Council--Polish KBN
joint research programme, and Polish KBN grants 2P03B 05119, 5P03B
14420.

\section*{Appendix: charm production in proton--air collisions}

Here we present a simple parameterization which may be used to
reproduce the inclusive cross section of $c$-quark production in
proton--air collisions, based on the GBW model (which yields the
continuous curves in Figs.~2--5). To $\pm5\%$ accuracy for
$0.05<x<0.6$ we have
\be x \frac{d\sigma}{dx}\raisebox{1.58ex}{$(p+{\rm air}\ra
c+\dots)$} = Ax^\beta(1-x^{1.2})^n \ee
where $\beta  =  0.05 - 0.016\ln(E/10~{\rm TeV})$  and
$$ \left. \begin{array}{r @{\quad = \quad}l}
n & 7.6 + 0.025 \ln(E/10~{\rm TeV}) \\
A & 140 + (11\ln(E/0.1\ {\rm TeV}))^{1.65}\ \mu{\rm b}
\end{array} \right\} \
{\rm for}\ 10^4<E<10^8\ \GeV, $$

$$ \left. \begin{array}{r @{\quad = \quad}l}
n & 7.6 + 0.012 \ln(E/10~{\rm TeV}) \\
A & 4100 + 245\ln(E/10^5\ {\rm TeV})~\mu{\rm b}
\end{array} \right\} \ {\rm for}\
10^8<E<10^{11}\ \GeV.$$
For $E=10^{12}\ \GeV$, the parameterization overestimates the
cross section by about 10\% for $x<0.2$.




\end{document}